\documentclass{article}

\usepackage[preprint]{neurips_2025}
\usepackage{threeparttable}
\usepackage{array}
\usepackage[utf8]{inputenc} 
\usepackage[T1]{fontenc}    
\usepackage{hyperref}       
\usepackage{url}            
\usepackage{booktabs}       
\usepackage{amsfonts}       
\usepackage{nicefrac}       
\usepackage{fontawesome}    
\usepackage{microtype}      
\usepackage{xcolor}         
\usepackage{natbib}
\usepackage{graphicx}
\usepackage{multirow} 
\usepackage{amssymb}
\usepackage{graphicx}
\usepackage{tabularray}
\UseTblrLibrary{booktabs}
\usepackage{listings}

\usepackage{makecell}
\usepackage{subcaption} 
\lstdefinestyle{datasetstyle}{
    basicstyle=\ttfamily\small,   
    frame=single,                 
    backgroundcolor=\color{gray!10}, 
    breaklines=true,              
    postbreak=\mbox{\textcolor{red}{$\hookrightarrow$}\space}, 
    showstringspaces=false,       
}
\usepackage{caption}

\usepackage[utf8]{inputenc} 
\usepackage[T1]{fontenc}    
\usepackage{hyperref}       
\usepackage{url}            
\usepackage{booktabs}       
\usepackage{amsfonts}       
\usepackage{nicefrac}       
\usepackage{microtype}      
\usepackage{xcolor}         
\usepackage{natbib}
\usepackage{graphicx}
\usepackage{multirow} 

\title{RAM-W600: A Multi-Task Wrist Dataset and Benchmark for Rheumatoid Arthritis}

\author{%
  \textbf{Songxiao Yang}$^1$\thanks{Equal Contribution.}\quad \textbf{Haolin Wang}$^{2\,*}$\quad \textbf{Yao Fu}$^2$\quad \textbf{Ye Tian}$^3$\quad \textbf{Tamotsu Kamishima}$^2$\\ \textbf{Masayuki Ikebe}$^2$\quad \textbf{Yafei Ou}$^1$\thanks{Corresponding Author (\texttt{ou.y.ac@m.titech.ac.jp, yafei.ou@riken.jp})}\quad \textbf{Masatoshi Okutomi}$^1$\\ \\
  $^1$ Institute of Science Tokyo, Tokyo, Japan \\
  $^2$ Hokkaido University, Sapporo, Japan \\
  $^3$ The University of Tokyo, Tokyo, Japan \\
}

\begin{document}

\maketitle

\begin{abstract}\label{sec:abs}
Rheumatoid arthritis (RA) is a common autoimmune disease that has been the focus of research in computer-aided diagnosis (CAD) and disease monitoring. In clinical settings, conventional radiography (CR) is widely used for the screening and evaluation of RA due to its low cost and accessibility. The wrist is a critical region for the diagnosis of RA. However, CAD research in this area remains limited, primarily due to the challenges in acquiring high-quality instance-level annotations. (i) The wrist comprises numerous small bones with narrow joint spaces, complex structures, and frequent overlaps, requiring detailed anatomical knowledge for accurate annotation. (ii) Disease progression in RA often leads to osteophyte, bone erosion (BE), and even bony ankylosis, which alter bone morphology and increase annotation difficulty, necessitating expertise in rheumatology.
This work presents a multi-task dataset for wrist bone in CR, including two tasks: (i) wrist bone instance segmentation and (ii) Sharp/van der Heijde (SvdH) BE scoring, which is the first public resource for wrist bone instance segmentation. This dataset comprises 1048 wrist conventional radiographs of 388 patients from six medical centers, with pixel-level instance segmentation annotations for 618 images and SvdH BE scores for 800 images. This dataset can potentially support a wide range of research tasks related to RA, including joint space narrowing (JSN) progression quantification, BE detection, bone deformity evaluation, and osteophyte detection. It may also be applied to other wrist-related tasks, such as carpal bone fracture localization.
We hope this dataset will significantly lower the barrier to research on wrist RA and accelerate progress in CAD research within the RA-related domain.
  \\
  \small \textbf{\mbox{\faGithub\hspace{.25em} Benchmark \& Code:}} \href{https://github.com/YSongxiao/RAM-W600}{github.com/YSongxiao/RAM-W600}\\
  \raisebox{-0.3\height}{\includegraphics[width=0.35cm]{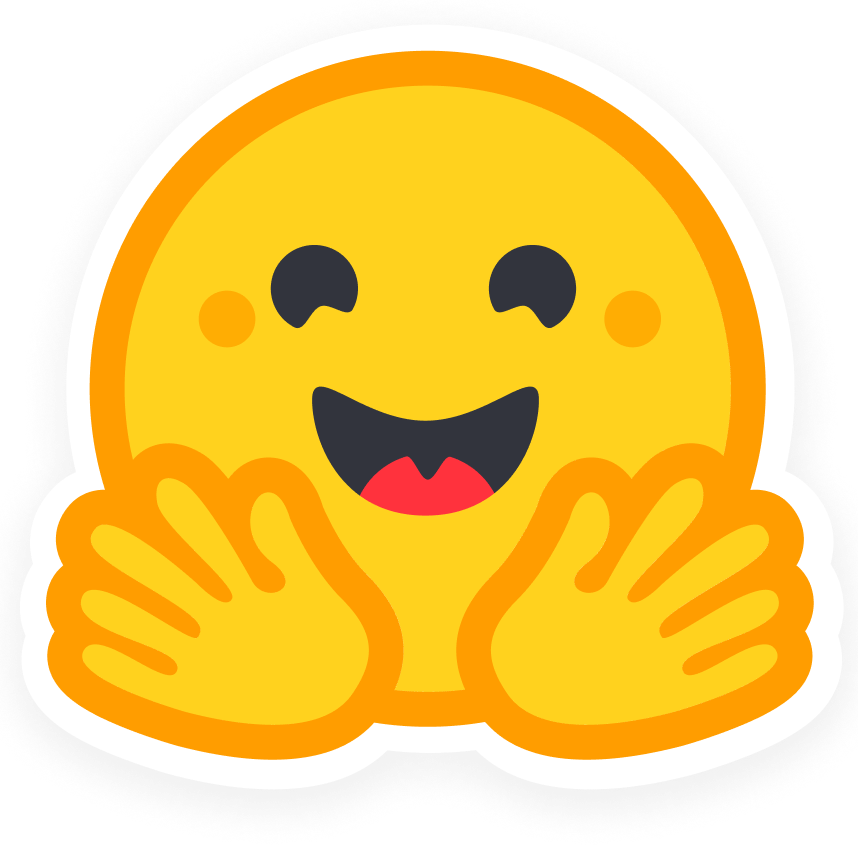}} \small \textbf{\mbox{Data \& Dataset Card:}} \href{https://huggingface.co/datasets/TokyoTechMagicYang/RAM-W600}{huggingface.co/datasets/TokyoTechMagicYang/RAM-W600}
\end{abstract}

\section{Introduction}


The wrist is a highly complex joint that facilitates a wide range of motion and bears substantial mechanical loads during daily activities. Due to its anatomical complexity and functional demands, the wrist is particularly susceptible to various pathological conditions~\citep{eschweiler2022anatomy}. Among these, rheumatoid arthritis (RA) is a common and debilitating autoimmune disease that frequently affects the wrist joint early in its progression~\citep{sharif2018rheumatoid}. It is marked by joint swelling and tenderness, which progressively leads to joint destruction and significant disability. Radiographic analysis plays a pivotal role in the diagnosis and management of RA, with joint space narrowing (JSN) progression and bone erosion (BE) serving as key markers for evaluating and tracking disease progression~\citep{aletaha2018diagnosis, ponnusamy2023automatic}. However, traditional radiographic assessment heavily relies on the radiologist’s expertise and subjective interpretation to detect subtle pathological features, which is time-consuming and often associated with limited accuracy and sensitivity. As a result, the development of computer-aided diagnostic (CAD) systems has attracted growing interest from both academic and industrial communities~\citep{stoel2024deep, kingsmore2021introduction, wang2025bls, wang2025layer}.

\begin{figure}[!t]
    \centering
    \includegraphics[width=\linewidth]{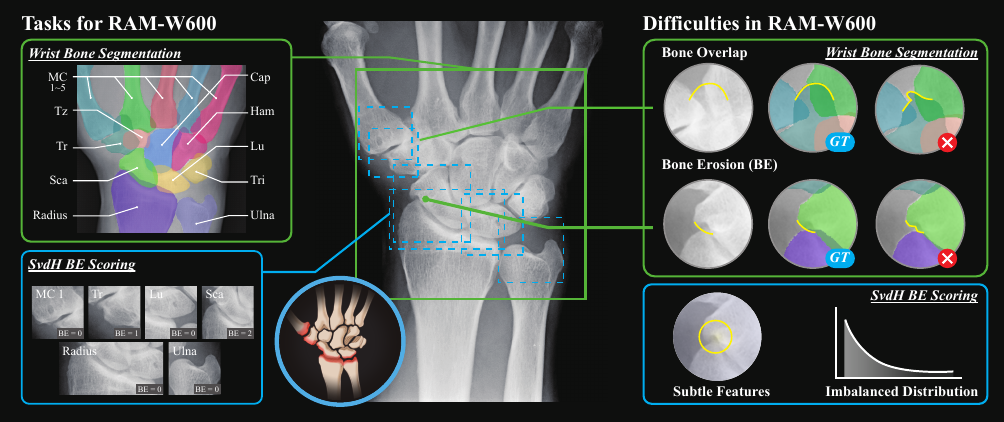}
    \caption{Overview of the RAM-W600 dataset, designed for wrist bone segmentation and SvdH BE scoring tasks. (MC 1 to 5: Metacarpal 1st to 5th; Tz: Trapezoid; Tr: Trapezium; Sca: Scaphoid; Radius: DistalRadius; Cap: Capitate; Ham: Hamate; Lu: Lunate; Tri: Pisiform $\&$ Triquetrum; Ulna: DistalUlna)}
    \label{fig:fig1}
\end{figure}

Accurate segmentation of wrist bones is critically important in medical image analysis, as it serves as a foundational step for numerous downstream tasks essential to the diagnosis and management of RA. These tasks include, but are not limited to, the evaluation of bone deformities, detection of osteophytes, and assessment of JSN. For example, in bone deformity analysis, precise segmentation is required to extract geometric features such as bone angles, alignment, and morphological irregularities across longitudinal scans~\cite{hirsiger2018corrective}. In osteophyte detection, segmented bone contours help identify abnormal bony outgrowths that are often hard to distinguish in raw radiographs due to anatomical overlap~\cite{overgaard2024artificial}. Similarly, accurate quantification of JSN depends on the precise boundaries between adjacent bones~\cite{huo2017automatic, ou2022sub, wang2023deep}. Segmentation errors can lead to incorrect inter-bone distance measurements, which are essential for monitoring disease progression.

However, the annotation process of a large-scale dataset is highly challenging and labor-intensive due to the anatomical complexity of the wrist and different pathological changes in the wrist bones. As shown in Fig.~\ref{fig:fig1}, 
(i) Obscured edges due to overlapping structures. The wrist, a structurally complex joint system, features tightly interlocked carpal bones~\cite{bhat2011radiographic}. This configuration frequently leads to overlapping phenomena in conventional radiography (CR), which significantly complicates the identification of each bone's outer edges. (ii) Morphological alterations resulting from pathological conditions. Due to the progression of RA and other pathological changes, BE, JSN, and osteophyte formation can affect certain bones or joints to varying degrees, often leading to substantial alterations in bone morphology~\cite{taouli2004rheumatoid,ezzati2022radiographic,hoving2004comparison}. Moreover, these factors may interact in diverse and combinatorial ways, further complicating the consistency and accuracy of annotations. 

Sharp/van der Heijde (SvdH) BE scoring~\cite{van2000read} is a widely recognized task in the automated diagnosis of RA. Nevertheless, it remains highly challenging due to difficulties in both annotation and model training. On the annotation side~\cite{raven2015radiocarpal,van1999reliability}, (i) accurate annotation demands specialized rheumatological expertise, as assessing the severity of BE is inherently complex. (ii) The process is subjective and prone to substantial inter-observer variability, resulting in inconsistent and uncertain ground truth labels. This subjectivity and ambiguity undermine the quality of supervision available for model training. From a training perspective, (i) the task is further complicated by a severe class imbalance, as cases of high-grade erosion are underrepresented in most clinical datasets. (ii) The pathological features of BE are often subtle, highly localized, small in scale, and demonstrate minimal variation across severity levels, thereby posing substantial challenges for automated detection and classification. Collectively, these factors render SvdH BE scoring a challenging task in developing robust and generalizable deep learning models for RA assessment.

In this paper, we introduce \textbf{R}heumatoid \textbf{A}rthritis \textbf{M}odeling-\textbf{W}rist 600 (RAM-W600), a multi-task dataset for wrist bone in conventional radiography. 
It comprises 1048 wrist conventional radiographs of 388 patients from six medical centers. 
Among them, 618 high-resolution wrist radiographs are provided with expert-verified instance-level annotations for wrist bone segmentation, along with 4800 SvdH BE scores. This dataset is expected to support a wide range of downstream tasks, such as anatomical structure localization, erosion progression analysis, and automated disease staging, thereby contributing to the broader advancement of computer-aided diagnosis in RA. Our primary contributions are threefold:
\begin{itemize}
    \item \textbf{First Multi-Task dataset for RA}: RAM-W600 is the first public large-scale dataset dedicated to both segmentation and SvdH BE scoring tasks, providing a valuable benchmark for developing and validating deep learning algorithms in conventional radiographs. Its multi-institutional composition ensures diversity in acquisition conditions, enhancing the generalizability of trained models.
    \item \textbf{High-quality annotations}: We provide high-quality pixel-level annotations of the wrist bones, including careful handling of overlapping region boundaries, and SvdH BE score in the region of interest (ROI).
    \item \textbf{Comprehensive benchmarks}: We present a benchmark for wrist bone instance segmentation and SvdH BE scoring, enabling standardized evaluation and comparison of algorithms for automated RA assessment.
\end{itemize}

\section{Related Works}

\begin{table*}[!t]
\centering
\caption{Comparison between RAM-W600 and the publicly available annotated datasets. Ann/Img: Annotations per image.}
\label{tab:dataset_compare}
\begin{threeparttable}
\resizebox{\textwidth}{!}{
\begin{tabular}{cccccccccc}
\toprule
\multirow{2.5}{*}{\textbf{Modality}} &
\multirow{2.5}{*}{\textbf{Dataset}} &
\multirow{2.5}{*}{\textbf{Year}} &
\multirow{2.5}{*}{\textbf{\makecell{Images\\(Ann/Img)}}} &
\multirow{2.5}{*}{\textbf{\makecell{Resolution\\(mm/pixel)}}} &
\multirow{2.5}{*}{\textbf{\makecell{Age\\(Mean$\pm$SD)}}} &
\multicolumn{2}{c}{\textbf{Tasks}} &
\multirow{2.5}{*}{\textbf{Purpose}} \\
\cmidrule(lr){7-8}
& & & & & & \textbf{Mask} & \textbf{Score} & \\
\midrule
\multirow{3}{*}{CR} & Halabi et al.~\cite{halabi2019rsna} & 2019 & 14236 (15) & - & 0.35 & \checkmark & & BAA \\
& Sun et al.~\cite{sun2022crowdsourcing} & 2022 & 674 (31) & - & - & & \checkmark & RA \\
& Ours (RAM-W600) & 2025 & 618 (15) + 800 (6) & 0.15$^*$ & 49.86$\pm$20.26 & \checkmark & \checkmark & RA \\ \midrule
CT & Moore et al.~\cite{moore2007digital} & 2007 & 30 (15) & - & 26.25$\pm$3.33 & \checkmark & & - \\
\bottomrule
\end{tabular}}
\begin{tablenotes}
\item \textbf{BAA}: Bone Age Assessment; \textbf{*}: Internal cohorts only.
\end{tablenotes}
\end{threeparttable}
\end{table*}

\subsection{Hand Radiographic Datasets}

Although hand radiographic data are relatively easy to acquire, the complex anatomical structure of the hand and the inherent limitations of current imaging techniques present significant challenges for accurate annotation. As shown in Table~\ref{tab:dataset_compare}, these challenges are further intensified in disease-specific applications, such as the diagnosis and monitoring of RA, where high-quality, expert-annotated datasets remain scarce. Earlier efforts produced computed tomography (CT) datasets with segmentation masks~\cite{moore2007digital}, but these were limited by small sample sizes. More recently, Halabi et al.~\cite{halabi2019rsna} released a large-scale CR dataset annotated with segmentation masks; however, its utility is confined to pediatric bone age assessment and limited to selected phalangeal regions. In contrast, RA-specific datasets such as that of Sun et al.~\cite{sun2022crowdsourcing} provide severity scores but lack pixel-wise annotations, thereby constraining their applicability to tasks requiring precise image segmentation.


\subsection{Wrist Bone Segmentation}


\begin{table*}[!t]
\centering
\caption{Summary of recent works on wrist segmentation. Ann/Img: Annotations per image.}
\label{tab:seg_studies}
\begin{threeparttable}
\resizebox{\textwidth}{!}{
\begin{tabular}{cccccccccc c}
\toprule
\multirow{2.5}{*}{\textbf{Modality}} & 
\multirow{2.5}{*}{\textbf{Works}} & 
\multirow{2.5}{*}{\textbf{Year}} & 
\multirow{2.5}{*}{\textbf{Backbone}} & 
\multirow{2.5}{*}{\textbf{Dataset}} & 
\multirow{2.5}{*}{\textbf{\makecell{Images\\(Ann/Img)}}} & 
\multirow{2.5}{*}{\textbf{\makecell{Age\\(Mean$\pm$SD)}}} & 
\multicolumn{3}{c}{\textbf{Objects}} & 
\multirow{2.5}{*}{\textbf{Purpose}} \\
\cmidrule(lr){8-10}
& & & & & & & \textbf{F} & \textbf{C} & \textbf{UR} & \\
\midrule
\multirow{4}{*}{CR} & Yang et al.~\cite{yang2021deep} & 2021 & ResNet & Private & 720 (2) & 36$\pm$13 &  &  & \checkmark & BMD \\
& Kang et al.~\cite{kang2022automatic} & 2022 & Mask R-CNN & Private & 702 (10) & - &  & \checkmark & \checkmark & - \\
& Lee et al.~\cite{lee2023osteoporosis} & 2023 & SAM & Private & 192 (7) & - & \checkmark &  & \checkmark & BMD \\
& Du et al.~\cite{du2024hand} & 2024 & GRU-Unet & \cite{halabi2019rsna} \& Private & 2000 (13) & - & \checkmark &  & \checkmark & BAA \\
\midrule
\multirow{2}{*}{CT} & Anas et al.~\cite{anas2016automatic} & 2016 & - & \cite{moore2007digital} \& Private & 60 (15) & - & \checkmark & \checkmark & \checkmark & - \\
& Sebro et al.~\cite{sebro2022machine} & 2022 & - & Private & 196 (17) & 64.9$\pm$8.7 & \checkmark & \checkmark & \checkmark & BMD \\
4DCT & Teule et al.~\cite{teule2024automatic} & 2024 & nnU-Net & Private & 19 (9) & - &  & \checkmark & \checkmark & - \\ \midrule
\multirow{4}{*}{MRI}  & Foster et al.~\cite{foster2018wrist} & 2018 & - & Private & 160 (8) & 47.1$\pm$9.25 &  & \checkmark &  & OA \\
& Radke et al.~\cite{radke2021deep} & 2021 & CNN & Private & 56 (8) & 30.7$\pm$13.6 &  & \checkmark & \checkmark & LWI \\
& Yiu et al.~\cite{yiu2024automated} & 2024 & nnU-Net & Private & 80 (15) & 54$\pm$12 & \checkmark & \checkmark & \checkmark & RA(BME) \\
& Raith et al.~\cite{raith2025multi} & 2025 & 3D U-Net & Private & 15 (8) & 27.8$\pm$3.11 &  & \checkmark &  & - \\
\bottomrule
\end{tabular}
}
\begin{tablenotes}
\footnotesize
\item \textbf{F}: Finger Bones; \textbf{C}: Carpal Bones; \textbf{UR}: Radius and Ulna Bones;\\
\textbf{BMD}: Bone Mineral Density; 
\textbf{BAA}: Bone Age Assessment; \\
\textbf{RA(BME)}: Rheumatoid Arthritis with Bone Marrow Edema;
\textbf{LWI}: Ligamentous Wrist Injuries.
\end{tablenotes}
\end{threeparttable}
\end{table*}


Radiological analysis of the wrist bones is central to the study of hand-related disorders. In particular, image segmentation plays an important role and holds significant value for both clinical practice and research, as summarized in Table~\ref{tab:seg_studies}. Notable progress has been made in wrist bone segmentation using various imaging modalities, including CT and magnetic resonance imaging (MRI). Early studies employed mathematical modeling techniques to achieve relatively mature segmentation outcomes on CT and MRI scans~\cite{anas2016automatic,foster2018wrist}. With recent advances in deep learning, both 2D and 3D segmentation of wrist bones in CT and MRI has further matured~\cite{yiu2024automated,teule2024automatic,radke2021deep,sebro2022machine,raith2025multi}, enabling more specialized investigations into disease-induced bone pathologies.
In contrast, research on wrist bone segmentation from radiographs is still limited. Although several deep learning-based methods have been proposed \cite{yang2021deep,kang2022automatic,du2024hand,lee2023osteoporosis}, few studies focus on complex pathological conditions such as RA. Due to the limitations of CR imaging, its two-dimensional nature causes anatomical overlap, tissue superposition, and low contrast, which make it difficult to identify bone boundaries and anatomical structures. In addition, although CR is more accessible and cost-effective than CT or MRI, accurate annotation is still difficult, especially in cases with active osteoarticular lesions. As a result, there are few high-quality, publicly available annotated datasets. This lack of data makes it hard to train and evaluate reliable segmentation models.

Consequently, achieving high-precision wrist bone segmentation in radiographs of patients with complex pathological conditions remains a critical challenge. Addressing this issue holds substantial potential for advancing efficient and user-friendly clinical decision support systems.


\begin{table*}[!t]
\centering
\caption{Summary of recent works on RA-related scoring. Ann/Img: Annotations per image.}
\label{tab:be_studies}
\begin{threeparttable}
\resizebox{\textwidth}{!}{
\begin{tabular}{cccccccccc}
\toprule
\multirow{2.5}{*}{\textbf{Modality}} & 
\multirow{2.5}{*}{\textbf{Works}} & 
\multirow{2.5}{*}{\textbf{Year}} & 
\multirow{2.5}{*}{\textbf{Backbone}} & 
\multirow{2.5}{*}{\textbf{Dataset}} & 
\multirow{2.5}{*}{\textbf{\makecell{Images\\(Ann/Img)}}} & 
\multirow{2.5}{*}{\textbf{Patients}} & 
\multirow{2.5}{*}{\textbf{\makecell{Age\\(Mean$\pm$SD)}}} & 
\multicolumn{2}{c}{\textbf{Tasks}} \\
\cmidrule(lr){9-10}
& & & & & & & & \textbf{SvdH BE} & \textbf{Others} \\
\midrule
\multirow{9}{*}{CR} & Hirano et al.~\cite{hiranodevelopment} & 2019 & CNN & Private & 216 (15) & 108 & 64.9$\pm$4.87 & \checkmark & SvdH JSN \\
& Ureten et al.~\cite{ureten2020detection} & 2020 & CNN & Private & 180 (2) & 180 & - &  & RA \& HC \\
& Maziarz et al.~\cite{maziarz2021deep} & 2021 & Unet & ~\cite{sun2022crowdsourcing} & 674 (31) & 562 & - & & Damage \\
& Hioki et al.~\cite{hioki2021evaluation} & 2021 & Yolo V3 & Private & 50 (4) & - & - & & Destruction \\
& Miyama et al.~\cite{miyama2022deep} & 2022 & DNN & Private & 226 (31) & 40 & 61.5$\pm$11.6 & \checkmark & SvdH JSN \\
& Wang et al.~\cite{wang2022deep} & 2022 & Yolo & Private & 915 (30) & 400 & $>$20 &  & mTSS \\
& Sun et al.~\cite{sun2022crowdsourcing} & 2022 & DNN & ~\cite{sun2022crowdsourcing} & 674 (31) & 562 & - & \checkmark & SvdH JSN \\
& Bo et al.~\cite{bo2024deep} & 2024 & ResNet & Private & 3818 (10) & - & - & \checkmark & SvdH JSN \\
& Lien et al.~\cite{lien2025deep} & 2025 & Yolo V7 & Private & 823 (30) & - & $>$20 &  & mTSS \\
\midrule
HR-pQCT & Folle et al.~\cite{folle2022deep} & 2022 & GradCAM & Private & 932 (3) & 617 & 45$\pm$15 & & HC \& RA \& PsA \\
MRI & Schlereth et al.~\cite{schlereth2024deep} & 2024 & CNN & Private & 211 (66) & 112 & 54.1$\pm$12.4 & \checkmark & osteitis \& synovitis \\
\bottomrule
\end{tabular}
}
\begin{tablenotes}
\footnotesize
\item \textbf{mTSS}: modified total Sharp Score; 
\textbf{PsA}: psoriatic arthritis; 
\textbf{HC}: healthy controls.
\end{tablenotes}
\end{threeparttable}
\end{table*}

\subsection{Detection and Assessment of BE}

The SvdH scoring system has been widely used to evaluate various joint abnormalities in RA. As summarized in Table~\ref{tab:be_studies}, an increasing number of automated methods have been developed in recent years to facilitate RA radiograph scoring. These approaches are typically based on the SvdH system and aim to assess key indicators such as JSN, BE, and the modified total Sharp score (mTSS). Most models are trained and validated on private datasets. Earlier studies primarily employed convolutional neural networks (CNNs) for feature extraction and classification~\cite{hirano2019development,miyama2022deep,sun2022crowdsourcing,bo2024deep}. Recently, object detection-based models have been introduced \cite{hioki2021evaluation,lien2025deep}, enabling the integration of lesion localization and scoring within end-to-end pipelines and enhancing both automation and usability. Some studies have explored RA classification and severity assessment using scoring systems other than SvdH~\cite{ureten2020detection,maziarz2021deep,hioki2021evaluation}.
Furthermore, research on automated RA assessment has expanded to encompass various imaging modalities, including MRI~\cite{schlereth2024deep} and high-resolution peripheral quantitative computed tomography (HR-pQCT)~\cite{folle2022deep}, along with the investigation of alternative scoring methods and evaluation standards, thereby further advancing the field of RA imaging analysis.

In summary, the wrist joint is one of the most anatomically complex and diagnostically significant regions in RA radiographs, offering substantial clinical and research value. Notably, the integration of precise wrist bone segmentation and lesion scoring within a multi-task learning framework has emerged as a key direction in advancing automated RA analysis. However, publicly available hand CR datasets remain significantly limited, particularly those focused on the wrist. Most datasets lack high-precision segmentation masks specifically annotated for the wrist region, and their corresponding BE scores are often incomplete or missing. This limits their suitability for RA-specific research, which requires high-quality, multi-dimensional annotated data. Therefore, the development of a wrist-focused CR dataset with detailed anatomical annotations and validated clinical scores is essential for the progress of intelligent RA imaging assessment.


\section{Overview of Dataset} \label{sec:ethic_review}

\paragraph{Ethical Considerations}
RAM-W600 dataset is in compliance with the guidelines of the Declaration of Helsinki and obtained approval from the Ethics Committee of Hokkaido University (approval number: 24-104) and Institute of Science Tokyo (approval number: A24672). All radiographs included in this dataset were collected with informed consent for research use and public release.

\subsection{Image and Annotation}
The dataset consisted of 1048 hand posteroanterior projection (PA) radiographs from 207 patients with RA and 181 patients without RA. The images were obtained from six different institutions: Hokkaido Medical Center for Rheumatic Diseases (HMCRD) (Sapporo, Japan), Sapporo City General Hospital (SCGH) (Sapporo, Japan), Hokkaido University (HU) (Sapporo, Japan), Digital Hand Atlas (DHA) from the University of Southern California (CA, US)~\cite{cao2000digital}, Bone Tumor X-ray Radiograh Dataset (BTXRD) from Monash University (Melbourne, Australia)~\cite{yao2025radiograph}, and FracAtlas (FA) from Islamic University of Technology (Gazipur, Bangladesh)~\cite{abedeen2023fracatlas}. Each institution has its own CR systems, and the dataset is managed using the digital imaging and communications in medicine (DICOM) standard, with the detailed information of imaging parameters referred to Table~\ref{tab:madical_info_AP}.

We employed specialized imaging processing methodologies to systematically construct wrist joint data. Initially, image cropping techniques were applied to focus on the wrist region, effectively eliminating interference from extraneous anatomical structures. 
Annotation was performed by a dedicated team consisting of a radiological technologist and two clinically experienced experts, including a board-certified radiologist with 25 years of experience and an orthopedic doctor with 5 years of clinical practice. This multidisciplinary expertise ensured that the annotations were both medically accurate and clinically relevant. For the segmentation task, initial contours were delineated by the radiological technologist and subsequently verified by the radiologist. For the classification task, three annotators independently assigned labels, and any discrepancies were resolved through discussion and consensus.
Based on this protocol, the annotation comprised three principal components:
\begin{itemize}
    \item \textbf{Anatomical Structure Annotation}: Precise contour delineation was performed for 14 wrist bones, including the first-fifth metacarpals (MC1-5), trapezium (Tr), trapezoid (Tz), scaphoid (Sca), lunate (Lu), capitate (Cap), hamate (Ham), pisiform $\&$ triquetrum (Tri), distal radius (Radius), and distal ulna (Ulna). A multi-label annotation strategy was implemented to independently mark each osseous structure.
    \item \textbf{Bone Location Annotation}: The SvdH BE scoring system focuses on five key joint regions: Metacarpal 1st, Trapezoid, Scaphoid, Lunate, Distal Radius, and Distal Ulna. We performed ROI annotations on these areas.
    \item \textbf{SvdH BE Scoring Annotation}: BE assessment was conducted using the SvdH scoring system, specifically targeting five critical articular groups: Metacarpal 1st, Trapezoid, Scaphoid, Lunate, Distal Radius, and Distal Ulna. This systematic evaluation focused on quantifying erosive changes at these predetermined anatomical sites.
\end{itemize}

With the division of these images, a comprehensive annotation pipeline was adopted, including professional annotators and strict inspection procedures. Further details of the data division and annotation can be found in Sec.~\ref{sec:detail_info}.



\begin{figure}[!t]
    \centering
    \includegraphics[width=1\linewidth]{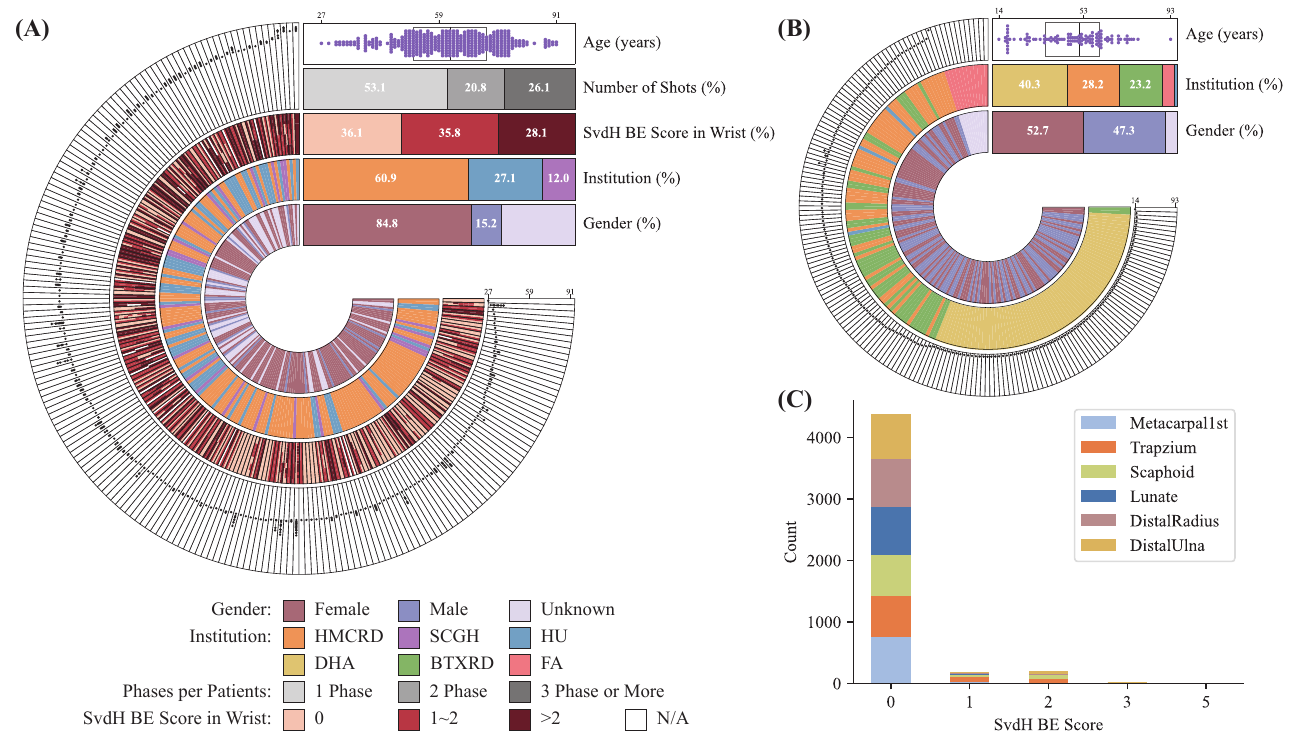}
    \caption{Distribution and Statistics for the age, gender, institution, number of shots, and BE scores in the RAM-W600 dataset. (A) Circular overview of the RA cohort. Each bar around the circular plot represents a unique patient. The concentric layers from inner to outer encode: (i) Gender distribution. (ii) Institution distribution. (iii) SvdH BE scores in both wrists for each study. Patients with multiple shots are represented multiple times in this layer. (iv) The patient’s age at each acquisition. (B) Circular overview of the Non-RA cohort. Similar to (A), each bar around the circular plot represents a unique patient. (C) Distribution of SvdH BE scores by joint surface.}
    \label{fig: data_distribution}
\end{figure}

\subsection{Statistics of RAM-W600}

We present statistical analyses of the RAM-W600 dataset to characterize both the RA cohort and the Non-RA cohort. Key attributes, including patient demographics (age, gender), institutional sources, follow-up frequency (phase), and BE scores, were systematically examined. In addition, joint-specific BE score distributions were compared across anatomical locations. Detailed statistics are summarized in Fig.~\ref{fig: data_distribution}. The RA cohort (A) collected from HMCRD, SCGH, and HU primarily consists of female patients. This pattern accords with epidemiologic evidence, since RA occurs most frequently in women between 30 and 50 years of age~\cite{pfeil2007computer,kvien2006epidemiological}. Most wrist joints in this cohort are annotated with an SvdH BE score of 0, indicating minimal erosive changes, while non-zero scores remain relatively uncommon. In addition, most patients underwent only a single imaging phase, and the age distribution spans a broad range. The Non-RA cohort (B) includes healthy controls from HMCRD and HU, as well as additional cohorts from DHA, BTXRD, and FA. This group exhibits a more balanced gender ratio and also shows a broad age distribution. Joint-level SvdH BE annotations in (C) reveal a highly imbalanced distribution across joint surfaces, with the vast majority of joint faces assigned a score of 0. Higher scores, such as 3 or 5, are nearly absent.
Such an imbalanced distribution has been commonly reported in clinical cohorts~\cite{bruynesteyn2002determination, jansen2001predictors}.
With advances in medical care, early detection and the effective use of disease-modifying treatments have markedly reduced the number of patients progressing to late-stage RA, making high BE scores increasingly rare in modern cohorts. Moreover, CR imaging is primarily performed to monitor early and moderate stages of RA, while advanced stages are less frequently imaged in current clinical practice.

\section{Experiments and Benchmarks}
\subsection{Wrist Bone Segmentation}
To evaluate wrist bone instance segmentation performance, we tested a series of widely used supervised architectures and their variants on the RAM-W600 dataset, as well as recent foundation models.
The supervised architectures included Unet~\cite{ronneberger2015u}, DeepLabV3~\cite{chen2017rethinking}, FPN~\cite{lin2017feature}, PSPNet~\cite{zhao2017pyramid}, DeepLabV3+~\cite{chen2018encoder}, SegResNet~\cite{myronenko20183d}, Unet++~\cite{zhou2018unet++}, SegFormer~\cite{xie2021segformer}, TransUNet~\cite{chen2021transunet}, UKAN~\cite{li2025u}, UMambaBot~\cite{ma2024u}, UMambaEnc~\cite{ma2024u}, SwinUMamba~\cite{liu2024swin} , while the foundation models comprised SAM~\cite{kirillov2023segment} and MedSAM~\cite{MedSAM}. In line with standard practice, segmentation performance was quantified using Dice Similarity Coefficient (DSC)~\cite{dice1945measures}; Normalized Surface Dice (NSD)~\cite{nikolov2021clinically}; Volumetric Overlap Error (VOE)~\cite{taha2015metrics}; Mean Surface Distance (MSD)~\cite{taha2015metrics}; and Relative Absolute Volume Difference (RAVD)~\cite{taha2015metrics}. The threshold for the NSD was set to 2 pixels.

\paragraph{Implementation details} \label{sec:implement_seg}
The dataset was split according to the configuration shown in Table~\ref{tab:dataset_distribution} (a) in Sec.~\ref{sec:data_pre}. BE and Non-BE cases were stratified using the SvdH BE score, where radiographs with a total BE score greater than 0 were considered BE cases. Cases were stratified based on the SvdH BE score, where radiographs with a total score greater than zero were classified as BE, while those with a score of zero were classified as non-BE.
For supervised models, all experiments were repeated five times on a single NVIDIA RTX 4090 GPU using five fixed random seeds (1024, 2025, 3407, 4096, and 5214) to ensure reproducibility, whereas foundation models were evaluated by a single inference run without repetition.
All radiographs were resized to 512$\times$512 pixels and used as input to the model. Model training employed the AdamW optimizer with a weight decay of 1e-2. The initial learning rate was set to 1e-4 and decayed according to a cosine annealing schedule (CosineAnnealingLR). Training was carried out for 100 epochs using a batch size of 8 and standard data augmentation techniques.

\begin{table*}[!t]
\centering
\caption{Instance segmentation results obtained on the Test set. The best results in each column are highlighted in \textbf{bold}, and the second-best values are \underline{underlined}.}
\begin{threeparttable}
\resizebox{\textwidth}{!}{
\setlength{\tabcolsep}{1pt} 
\begin{tabular}{l *{2}{ccl}cccc}
\toprule
\multirow{2.5}{*}{\textbf{Model}} & \multicolumn{3}{c}{\textbf{DSC $\uparrow$ (\%)}} & \multicolumn{3}{c}{\textbf{NSD $\uparrow$ (\%)}} & \multirow{2.5}{*}{\textbf{\makecell{VOE $\downarrow$\\(\%)}}} & \multirow{2.5}{*}{\textbf{\makecell{MSD $\downarrow$\\(pix)}}} & \multirow{2.5}{*}{\textbf{\makecell{Params\\(M)}}} & \multirow{2.5}{*}{\textbf{\makecell{Time\\(ms)}}}\\
\cmidrule(lr){2-4}
\cmidrule(lr){5-7}
& \textbf{BE} & \textbf{nonBE} & \multicolumn{1}{c}{\textbf{All}} & \textbf{BE} & \textbf{nonBE} & \multicolumn{1}{c}{\textbf{All}} & & & & \\
\midrule
\multicolumn{11}{c}{\textbf{Supervised Models}}  \\
\midrule
Unet~\cite{
ronneberger2015u}           & 96.70$\pm$0.05 & 96.83$\pm$0.09 & 96.79$\pm$0.08 & 83.59$\pm$0.41 & 83.27$\pm$0.58 & 83.36$\pm$0.52 & 6.13$\pm$0.14 & 1.83$\pm$0.10 & 7.94 & 13.57 \\
DeepLabV3~\cite{chen2017rethinking}    & 96.55$\pm$0.03 & 96.86$\pm$0.02 & 96.78$\pm$0.02*** & 82.17$\pm$0.28 & 82.89$\pm$0.23 & 82.69$\pm$0.19* & 6.20$\pm$0.03 & 1.37$\pm$0.01 & 26.00 & 9.19 \\
FPN~\cite{lin2017feature}              & 96.59$\pm$0.07 & 96.85$\pm$0.07 & 96.78$\pm$0.07*** & 81.45$\pm$0.68 & 81.83$\pm$0.63 & 81.73$\pm$0.64 & 6.19$\pm$0.13 & 1.38$\pm$0.02 & 23.15 & 8.43 \\
PSPNet~\cite{zhao2017pyramid}          & 95.30$\pm$0.05 & 95.55$\pm$0.07 & 95.48$\pm$0.06* & 71.58$\pm$0.46 & 71.02$\pm$0.47 & 71.17$\pm$0.45 & 8.52$\pm$0.10 & 2.05$\pm$0.04 & 21.49 & 4.46 \\
DeepLabV3+~\cite{chen2018encoder}      & 96.78$\pm$0.01 & 97.01$\pm$0.03 & 96.95$\pm$0.02*** & 83.56$\pm$0.17 & 83.73$\pm$0.27 & 83.68$\pm$0.20 & 5.87$\pm$0.04 & 1.31$\pm$0.02 & 22.43 & 5.57 \\
SegResNet~\cite{myronenko20183d}       & 96.48$\pm$0.21 & 96.64$\pm$0.20 & 96.60$\pm$0.20* & 81.78$\pm$1.25 & 81.79$\pm$1.14 & 81.79$\pm$1.17 & 6.50$\pm$0.37 & 1.79$\pm$0.21 & 1.60 & 4.93 \\
Unet++~\cite{zhou2018unet++}           & 97.21$\pm$0.02 & 97.37$\pm$0.04 & 97.33$\pm$0.03* & 86.85$\pm$0.26 & 87.04$\pm$0.26 & 86.99$\pm$0.23 & 5.15$\pm$0.06 & 1.36$\pm$0.07 & 2.41 & 14.83 \\
SegFormer~\cite{xie2021segformer}      & 96.82$\pm$0.06 & 97.09$\pm$0.02 & 97.01$\pm$0.03*** & 84.24$\pm$0.46 & 84.65$\pm$0.20 & 84.53$\pm$0.25 & 5.74$\pm$0.06 & 1.28$\pm$0.00 & 21.87 & 5.04 \\
TransUNet~\cite{chen2021transunet}     & \underline{97.50$\pm$0.04} & \underline{97.67$\pm$0.06} & \underline{97.62$\pm$0.05***} & \underline{89.20$\pm$0.24} & \underline{89.59$\pm$0.36} & \underline{89.48$\pm$0.33} & \underline{4.60$\pm$0.10} & \underline{1.05$\pm$0.03} & 105.91 & 22.05 \\
UKAN~\cite{li2025u}                    & 96.74$\pm$0.06 & 96.98$\pm$0.05 & 96.91$\pm$0.05*** & 83.15$\pm$0.22 & 83.41$\pm$0.16 & 83.33$\pm$0.16 & 5.93$\pm$0.10 & 1.34$\pm$0.04 & 6.36 & 10.30 \\
UMambaBot~\cite{ma2024u}               & 97.40$\pm$0.04 & 97.58$\pm$0.02 & 97.53$\pm$0.03** & 88.77$\pm$0.23 & 88.94$\pm$0.18 & 88.89$\pm$0.20 & 4.76$\pm$0.05 & 1.13$\pm$0.01 & 4.42 & 15.12 \\
UMambaEnc~\cite{ma2024u}               & 97.44$\pm$0.05 & 97.61$\pm$0.03 & 97.56$\pm$0.03** & 88.92$\pm$0.31 & 89.17$\pm$0.29 & 89.10$\pm$0.28 & 4.71$\pm$0.06 & 1.11$\pm$0.02 & 4.58 & 16.44 \\
SwinUMamba~\cite{liu2024swin}          & \textbf{97.65$\pm$0.02} & \textbf{97.80$\pm$0.02} & \textbf{97.75$\pm$0.02**} & \textbf{90.56$\pm$0.12} & \textbf{90.77$\pm$0.15} & \textbf{90.71$\pm$0.14} & \textbf{4.35$\pm$0.03} & \textbf{1.06$\pm$0.05} & 59.89 & 38.52 \\
\midrule
\multicolumn{11}{c}{\textbf{Foundation Models}}  \\
\midrule
SAM (box)~\cite{kirillov2023segment}                              & 88.91$\pm$5.59 & 88.67$\pm$4.80 & 88.74$\pm$5.01 & 65.91$\pm$6.06 & 63.82$\pm$7.32 & 64.40$\pm$7.03 & 18.45$\pm$5.23 & 4.25$\pm$1.46 & 641.09 & 193.47 \\
SAM (pt)~\cite{kirillov2023segment}                               & 80.18$\pm$7.10 & 80.46$\pm$11.13 & 80.38$\pm$10.14 & 55.56$\pm$9.93 & 55.84$\pm$12.06 & 55.76$\pm$11.47 & 28.42$\pm$10.82 & 18.21$\pm$16.08 & 641.09 & 32.72 \\
MedSAM (box)~\cite{MedSAM}                           & 85.07$\pm$2.05 & 85.06$\pm$2.69 & 85.07$\pm$2.52 & 39.91$\pm$6.46 & 38.38$\pm$7.28 & 38.81$\pm$7.07 & 25.15$\pm$3.59 & 5.97$\pm$1.19 & 93.74 & 99.48 \\
\bottomrule
\end{tabular}
}
\raggedright
\begin{tablenotes}
\item Time: Inference time per image on RTX 4090 GPU.
\item Foundation models: one inference (mean $\pm$ std across cases).\newline Supervised models: five runs (mean $\pm$ std across runs).
\item Mann-Whitney U Test between BE $\&$ nonBE, *: P < 0.05; **: P < 0.01; ***: P < 0.001.
\end{tablenotes}
\end{threeparttable}
\label{seg_results}
\end{table*}

\paragraph{Benchmark results}
The results shown in Table~\ref{seg_results} demonstrate that mainstream supervised models achieve outstanding performance in terms of DSC, with the highest value reaching 97.75\% (SwinUMamba), indicating robust overlap accuracy in global segmentation regions. However, NSD values remain comparatively low (peak: 90.71\%), with significant variations across models, highlighting persistent challenges in bone boundary delineation. This limitation is closely tied to the inherent complexities of wrist bone segmentation: inter-bone occlusions leading to blurred boundaries, and BE regions characterized by abnormal texture and edge variations, which further exacerbate segmentation difficulty. Meanwhile, group analysis reveals statistically significant differences (p < 0.05–0.001) in DSC between BE and nonBE samples for most models, confirming the detrimental impact of BE on segmentation performance. In contrast, the NSD metric exhibited no statistically significant differences between groups. This discrepancy may stem from the heightened sensitivity of NSD to boundary errors and the larger variance in boundary-related discrepancies within the dataset, underscoring the intrinsic difficulty in handling bone edges. In addition, foundation models such as SAM and MedSAM achieved lower DSC ($\leq$ 88.7\% for SAM and 85.1\% for MedSAM) and NSD ($\leq$ 75.0\%) compared with supervised models, further demonstrating the limited adaptability of general-purpose segmentation priors to the specialized task of wrist bone delineation. In conclusion, the primary bottleneck in this task lies in improving model robustness for complex bone boundaries.


\begin{figure}[!t]
    \centering
    \includegraphics[width=1\linewidth]{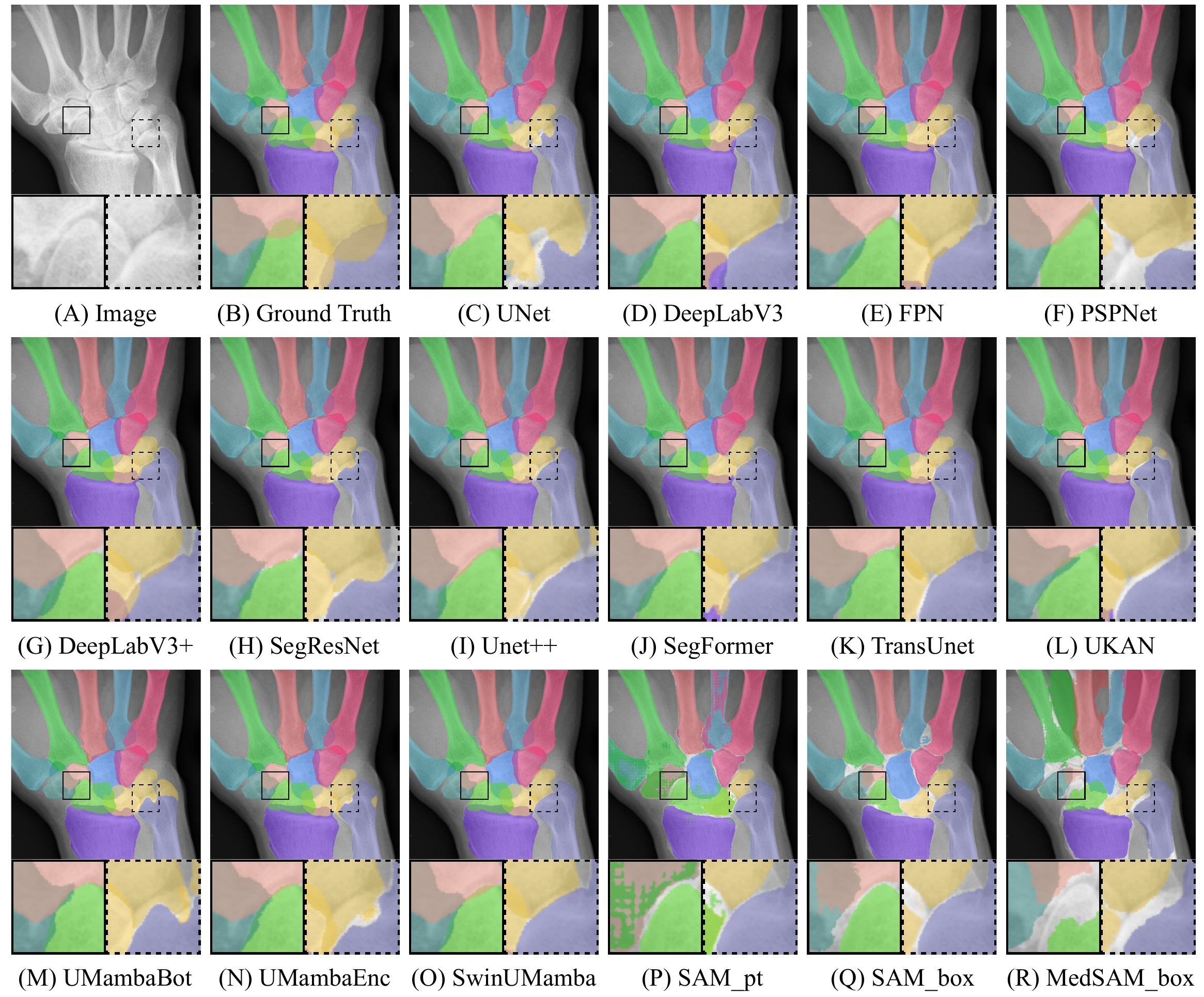}
    \caption{Wrist bone segmentation visualization results. The solid box indicates segmentation challenges caused by BE, while the dashed box represents difficulties arising from bone overlap.}
    \label{fig: seg_resutls}
\end{figure}

\paragraph{Visualization}
Some representative results are shown in Fig. \ref{fig: seg_resutls}. Compared to the ground truth, mainstream supervised networks exhibit performance degradation in segmenting bone edges with multi-layer occlusions, a challenge that becomes particularly pronounced under complex occlusion scenarios. Current models also demonstrate notable inconsistency, lacking reliable solutions to address this issue effectively. Furthermore, in the context of BE (RA), most existing architectures fail to adequately capture the inward collapse of bone edges caused by erosive changes. However, networks incorporating Mamba-based architectures show partial improvements in handling such morphological distortions, as evidenced by comparative analyses. Visualization results further corroborate the persistent challenges in this segmentation task, primarily attributed to bone overlaps and erosion-induced structural anomalies. These factors collectively lead to fragmented or inaccurate edge predictions, emphasizing the need for dedicated architectural innovations. In contrast, foundation models such as SAM and MedSAM exhibit less precise boundary localization and frequent edge discontinuities, underscoring their limited adaptability to the fine-grained requirements of wrist bone segmentation.


Unlike natural images or other medical imaging modalities such as MRI and CT, CR captures the cumulative attenuation of X-rays along their path, resulting in grayscale representations of internal structures. This often leads to overlapping anatomical features and blurred boundaries in two-dimensional images. 
Moreover, pathological BE caused by RA can induce notable morphological changes in bone structure, further complicating segmentation. Traditional image processing and segmentation techniques often struggle to accurately delineate overlapping bone boundaries or detect morphological abnormalities resulting from pathological alterations.
To address these challenges, future research may benefit from exploring multi-scale feature fusion strategies and advanced edge refinement techniques. Given the relatively fixed spatial arrangement of bones, incorporating global contextual information could be particularly advantageous for improving segmentation accuracy.

\subsection{Classification of BE}
The advanced binary classification methods of BE were evaluated on the RAM-W600 dataset. The selected classification models included MobileViT~\cite{mehta2021mobilevit}, ResNet~\cite{he2016deep}, MobileNet~\cite{howard2017mobilenets}, LeViT~\cite{graham2021levit}, EfficientFormer~\cite{li2022efficientformer}, MedMamba~\cite{yue2024medmamba}, and ConvKAN~\cite{bodner2024convolutional}. In line with standard practice, classification performance was quantified using balanced accuracy (BACC)~\cite{brodersen2010balanced}, F1-score~\cite{chinchor1993muc}, diagnostic odds ratio (DOR)~\cite{glas2003diagnostic}, accuracy (ACC)~\cite{hastie2009elements}, sensitivity (SEN)~\cite{altman1994diagnostic}, specificity (SPC)~\cite{altman1994diagnostic}, and precision (PRE)~\cite{sokolova2009systematic}.

\paragraph{Implementation details} \label{sec:implement_cls}
The dataset was split according to the configuration shown in Table~\ref{tab:dataset_distribution}~(b) in Sec.~\ref{sec:data_pre}. BE classification was performed on a joint-surface basis, focusing on the six joint surfaces of clinical interest. A joint surface was labeled as BE if its corresponding SvdH BE score was greater than 0. All experiments were repeated five times on a single NVIDIA RTX4090 GPU using five fixed random seeds (1024, 2025, 3407, 4096, 5214) to ensure reproducibility. 
All ROIs were resized to 224$\times$224 pixels and used as input to the model. Model training utilized the AdamW optimizer with a weight decay of 1e-2. The initial learning rate was set to 1e-6 and decayed using a cosine annealing schedule (CosineAnnealingLR).
Training was performed for 100 epochs with a batch size of 16 and standard data augmentation techniques.

\paragraph{Benchmark results}

\begin{table*}[!t]
\centering
\caption{BE $\&$ nonBE classification results obtained on the Test set. The best results in each column are highlighted in \textbf{bold}, and the second-best values are \underline{underlined}.}
\label{tab: be_results}
\begin{threeparttable}
\setlength{\tabcolsep}{2pt} 
\resizebox{\textwidth}{!}{
\centering
\begin{tabular}{lccccccccc}
\toprule
\textbf{Model} & \textbf{\makecell{BACC$\uparrow$\\(\%)}} & \textbf{\makecell{F1-Score$\uparrow$\\(\%)}} & \textbf{DOR$\uparrow$} & \textbf{\makecell{ACC$\uparrow$\\(\%)}} & \textbf{\makecell{SEN$\uparrow$\\(\%)}} & \textbf{\makecell{SPC$\uparrow$\\(\%)}} & \textbf{\makecell{PRE$\uparrow$\\(\%)}} & \textbf{Params} & \textbf{\makecell{Time\\(ms)}} \\
\midrule
MobileViT~\cite{mehta2021mobilevit}       & \textbf{52.64$\pm$0.61} & \underline{11.85$\pm$0.48} & \underline{1.82$\pm$0.19} & 81.42$\pm$0.87 & 21.06$\pm$0.93 & 84.23$\pm$1.13 & \underline{9.31$\pm$0.76} & 4.94M & 4.53 \\
ResNet~\cite{he2016deep}                  & \underline{51.75$\pm$1.02} & 10.89$\pm$1.06 & 1.16$\pm$0.41 & 78.27$\pm$1.31 & \underline{23.10$\pm$2.54} & 80.40$\pm$1.60 & 7.79$\pm$0.74 & 0.70M & 1.99 \\
MobileNet~\cite{howard2017mobilenets}     & 47.84$\pm$2.52 & 10.79$\pm$1.98 & 0.89$\pm$0.38 & 74.08$\pm$6.31 & 17.02$\pm$4.60 & 78.66$\pm$7.31 & 9.07$\pm$2.99 & 0.69M & 1.72 \\
LeViT~\cite{graham2021levit}              & 49.29$\pm$0.69 & 6.73$\pm$1.90 & 1.51$\pm$1.45 & 84.17$\pm$2.46 & 8.49$\pm$3.57 & 90.09$\pm$3.73 & 8.99$\pm$5.68 & 7.01M & 2.65 \\
EfficientFormer~\cite{li2022efficientformer} & 50.63$\pm$1.86 & \textbf{12.40$\pm$2.43} & 1.06$\pm$0.31 & 72.04$\pm$3.45 & \textbf{27.90$\pm$8.73} & 73.37$\pm$5.23 & 8.82$\pm$0.63 & 3.25M & 3.63 \\
MedMamba~\cite{yue2024medmamba}           & 50.83$\pm$1.00 & 6.91$\pm$3.51 & \textbf{5.89$\pm$9.66} & \underline{86.56$\pm$4.48} & 8.94$\pm$7.45 & \underline{92.73$\pm$6.55} & \textbf{11.56$\pm$7.98} & 14.45M & 6.06 \\
ConvKAN~\cite{bodner2024convolutional}    & 49.26$\pm$0.84 & 3.49$\pm$3.13 & 0.44$\pm$0.37 & \textbf{87.42$\pm$4.55} & 3.82$\pm$4.89 & \textbf{94.70$\pm$6.32} & 6.56$\pm$7.09 & 3.49M & 29.96 \\
\bottomrule
\end{tabular}
}
\raggedright
\begin{tablenotes}
\item Time: Inference time per image on RTX 4090 GPU.
\end{tablenotes}
\end{threeparttable}
\end{table*}

\begin{figure}[!t]
    \centering
    \includegraphics[width=\linewidth]{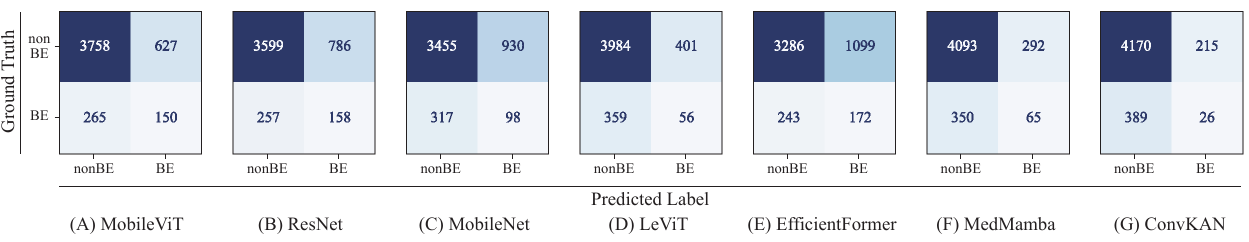}
    \caption{BE $\&$ nonBE confusion matrix results for classification of BE.}
    \label{fig: be_results}
\end{figure}

The results in Table~\ref{tab: be_results} reveal that mainstream models achieve only modest performance in terms of BACC and F1-score, with the best results reaching 52.64\% (MobileViT) and 12.40\% (EfficientFormer), respectively, indicating limited robustness in distinguishing BE from nonBE cases. In contrast, the DOR exhibits considerable variability across models, peaking at 5.89 (MedMamba). Notably, some models (e.g., ConvKAN) achieve relatively high specificity (94.70\%) while suffering from extremely low sensitivity (3.82\%), reflecting a strong bias toward negative predictions. This inconsistency across metrics underscores the difficulty of the task, likely stemming from extreme class imbalance and the subtle radiographic presentation of BE. The confusion matrices in Fig.~\ref{fig: be_results} further illustrate this imbalance, showing that all models consistently perform better on the majority class (nonBE) than on the minority class (BE), highlighting the inherent challenge of detecting subtle BE features.

Future research should further focus on enhancing the model's ability to detect subtle BE features under highly imbalanced data conditions. In clinical practice, early or mild BE lesions typically exhibit low visibility, presenting as small and inconspicuous regions that are easily confounded by overlapping bones, imaging artifacts, or noise. Although advanced BE lesions are more prominent in size, they often co-occur with other RA manifestations such as joint space narrowing and osteophyte formation, introducing additional sources of interference. These challenges collectively complicate the end-to-end scoring process for BE across different stages of the disease. To improve model performance on such difficult samples, future efforts may explore targeted augmentation strategies for minority classes or develop architectures capable of extracting weak pathological signals. Such advancements would enhance both the sensitivity and robustness of RA imaging assessment tools, thereby promoting their clinical applicability and translational value.




\section{Conclusions and Limitations}
We have introduced RAM-W600, the first publicly available multi-task CR dataset for RA assessment, which encompasses two key tasks: wrist bone segmentation and SvdH BE localization and scoring. RAM-W600 has provided high-quality pixel-level annotations for the anatomically complex wrist region, which often presents with severe bone overlapping and erosive changes. In addition to detailed annotations, the dataset includes benchmark results for both segmentation and BE scoring tasks. Experimental findings have demonstrated the considerable challenges posed by these tasks, including the accurate delineation of bones in the presence of occlusion and erosion in the segmentation task, and the robust scoring of affected joints in the grading task. By establishing RAM-W600 and its associated benchmarks, we have offered a valuable resource for advancing research in medical image analysis. This dataset has opened new avenues for the development and validation of robust CAD systems and holds promise for improving diagnostic accuracy and clinical decision-making in the management of RA.


Despite its contributions to advancing CAD for RA, the RAM-W600 dataset has several limitations. First, the RA cases are primarily derived from a single geographic region and a relatively homogeneous ethnic population, which may limit the generalizability of models trained on the dataset to more diverse clinical settings. This lack of demographic variability could reduce the robustness of model performance across different populations. Second, the distribution of SvdH BE scores is imbalanced, with certain score levels notably underrepresented. This imbalance poses challenges for both training and evaluation, particularly in learning fine-grained disease severity and ensuring consistent performance across all stages of RA progression.


For wrist bone segmentation, future research should focus on developing dedicated network architectures that incorporate multi-scale contextual information and boundary-sensitive mechanisms. Such designs are essential to address the challenges posed by anatomical complexity and projection-induced overlap in wrist radiographs, particularly for achieving accurate delineation in regions affected by bone overlap and BE. Regarding the SvdH BE scoring task, early-stage lesions often present weak radiographic signals and are obscured by overlapping structures, while advanced-stage cases commonly exhibit coexisting RA-related features, resulting in complex local characteristics. In addition, the highly imbalanced distribution of BE samples continues to hinder lesion recognition in current approaches. To overcome these limitations, it is crucial to design model components capable of extracting subtle pathological features, thereby improving sensitivity and robustness in detecting early-stage BE. Advancements in these directions are expected to significantly enhance the automation of RA wrist image analysis and reinforce its clinical utility in diagnosis and longitudinal disease monitoring.



\section*{Acknowledgments}
This work was supported by JST BOOST and JST SPRING, Japan Grant Number JPMJBS2426, JPMJSP2106 and JPMJSP2180.

\newpage
\appendix

\section{RAM-W600 Data Access and Format} \label{sec:data_access}

The data can be accessed on HuggingFace at \url{https://huggingface.co/datasets/TokyoTechMagicYang/RAM-W600}. The dataset has a permanent DOI: \url{https://doi.org//10.57967/hf/5328}. The benchmark and code can be accessed on Github at \url{https://github.com/YSongxiao/RAM-W600}.

The dataset is organised in two main folders (\texttt{Segmentation/} and \texttt{BE\_SvdH\_Prediction/}) corresponding to two tasks. The dataset structure is shown as follows:
\begin{lstlisting}[style=datasetstyle]
RAM-W600/
|-- JointLocationDetection/
|   |-- images/              # Contains all input images in BMP format
|       |-- 0145_0004_L.bmp
|       |-- 0145_0004_R.bmp
|       |-- ...
|   |-- Joints.coco.json     # Ground-truth annotations for joints' locations
|-- BoneSegmentation/
|   |-- images/                  # Contains all input images in BMP format
|       |-- 0001_0001_L.bmp              
|       |-- 0001_0001_R.bmp
|       |-- ...     
|   |-- masks/                   # Contains corresponding masks in NumPy (.npy) format
|       |-- train/
|           |-- 0006_0001_L.npy
|           |-- ...
|       |-- val/
|           |-- 0001_0001_R.npy
|           |-- ...
|       |-- test/
|           |-- 0002_0001_L.npy
|           |-- ...
|-- SvdHBEScoreClassification/
|   |-- train/
|       |-- 0003_0001_L/
|           |-- DistalRadius.bmp
|           |-- DistalUlna.bmp
|           |-- ...
|       |-- ...
|   |-- val/
|       |-- 0001_0001_R/
|           |-- DistalRadius.bmp
|           |-- DistalUlna.bmp
|           |-- ...
|       |-- ...
|   |-- test/
|       |-- 0005_0001_L/
|           |-- DistalRadius.bmp
|           |-- DistalUlna.bmp
|           |-- ...
|       |-- ...
|   |-- JointBE_SvdH_GT.json     # Ground-truth annotations for joint BE scores
|-- Metadata.xlsx       # Metadata for the dataset
            
\end{lstlisting}

\begin{itemize}
    \item \texttt{BoneSegmentation/images/}: Contains all original images in BMP format. Each file is named as \texttt{[PatientID]\_[StudyID]\_[L/R].bmp}, where L and R indicate the left or right hand, respectively. 
    \item \texttt{BoneSegmentation/masks/}: Contains the corresponding segmentation masks stored as NumPy arrays (.npy). The masks are organized into \texttt{train/}, \texttt{val/}, and \texttt{test/} subsets, with filenames matching the corresponding images.
    \item \texttt{JointLocationDetection/images/}: Contains all original images in BMP format. Each file is named as \texttt{[PatientID]\_[StudyID]\_[L/R].bmp}, where L and R indicate the left or right hand, respectively.
    \item \texttt{JointLocationDetection/Joints.coco.json}: A JSON file containing ground-truth annotations for the joint scores, indexed by case identifiers. The format of entries in JSON file is shown as follows:
    \begin{samepage}
        \begin{lstlisting}[style=datasetstyle]
{
  "images": [
    {
      "id": 0,
      "file_name": "0334_0001_R.bmp",
      "width": 600,
      "height": 600
    },
    ...
  ],
  "annotations": [
    {
      "id": 3281,
      "image_id": 546,
      "category_id": 1,
      "bbox": [170.0, 305.88, 235, 235],
      "area": 23680.95,
      "segmentation": [],
      "iscrowd": 0
    },
    ...
  ],
  "categories": [
    {
      "id": 1,
      "name": "DistalRadius",
      "supercategory": "joint"
    },
    {
      "id": 2,
      "name": "DistalUlna",
      "supercategory": "joint"
    },
    ...
  ]
}
        \end{lstlisting}
    \end{samepage}
    Each entry in the \texttt{images} list represents a wrist radiograph, while the \texttt{annotations} list contains bounding box annotations for individual joints, identified by their \texttt{category\_id}. The \texttt{categories} section maps category IDs to specific joint names such as \texttt{Lunate}, \texttt{Scaphoid}, and \texttt{Trapezium}.
    \item \texttt{SvdHBEScoreClassification/train/val/test/}: Each subset contains folders named as \texttt{[PatientID]\_[StudyID]\_[L/R]}, representing individual cases. Inside each folder are six ROI images in BMP format, each corresponding to different joint surfaces.
    \item \texttt{SvdHBEScoreClassification/JointBE\_SvdH\_GT.json}: A JSON file containing ground-truth annotations for the joint scores, indexed by case identifiers. The format of entries in JSON file is shown as follows:
    \begin{samepage}
    \begin{lstlisting}[style=datasetstyle]
{
    "identifier": "0035_0001_L",
    "patient_id": "0035",
    "study_id": "0001",
    "hand": "L",
    "joints": {
      "Metacarpal1st": 0,
      "Trapezium": 0,
      "Scaphoid": 0,
      "Lunate": 0,
      "DistalRadius": 0,
      "DistalUlna": 0
    }
}
    \end{lstlisting}
    \end{samepage}
    \item \texttt{Metadata.xlsx}: An Excel file containing patient-, study-, and image-level metadata. 
    It provides identifiers, demographic attributes, institutional sources, imaging parameters, and clinical reference scores. 
    The key columns are described as follows:
    \begin{itemize}
        \item \texttt{Mapped Image Stem}: A normalized identifier of each radiographic study in the format \texttt{XXXX\_XXXX}. 
        This stem represents the study itself rather than a direct image file. 
        The corresponding radiographs are determined by appending the hand side (\texttt{\_L} or \texttt{\_R}) to the stem, 
        which specifies the left or right hand image.
        \item \texttt{PatientID}: An anonymized patient identifier, allowing multiple studies from the same individual to be grouped.
        \item \texttt{StudyID}: An anonymized study identifier, denoting examinations at different time points.
        \item \texttt{IsRA}: Binary flag for rheumatoid arthritis status (1 = RA patient, 0 = non-RA control).
        \item \texttt{PatientSex}: Patient sex, recorded as \texttt{M} (male), \texttt{F} (female) or \texttt{O} (unknown).
        \item \texttt{PatientAge}: Age at the time of the study, expressed in years (e.g., 59.5).
        \item \texttt{InstitutionName}: Source institution where the radiograph was acquired (e.g., HMCRD, SCGH, HU).
        \item \texttt{StudyDate (Days)}: Relative day of the study, with baseline examination set to 0.
        \item \texttt{ImagerPixelSpacing}: In-plane resolution of the image in millimeters, recorded as \texttt{[row spacing, column spacing]}.
        \item \texttt{[Rows, Columns]}: Image resolution in pixels.
        \item \texttt{L / R}: Indicators for whether valid SvdH scores are available for the left or right hand (1 = available, 0 = unavailable).
        \item \texttt{SvdH\_L / SvdH\_R}: Total Sharp/van der Heijde erosion scores for the left and right hands.
        \item \texttt{Joint-specific scores}: Integer scores for six anatomical regions 
        (Metacarpal1st, Trapezium, Scaphoid, Lunate, DistalRadius, DistalUlna), recorded separately for left (\_L) and right (\_R) hands. 
        Higher scores indicate more severe erosion. 
    \end{itemize}
\end{itemize}

\section{Detailed Information of RAM-W600}\label{sec:detail_info}
\subsection{License and Attribution}\label{sec:license}
The conventional radiographs and associated annotations (segmentation masks and SvdH BE scores) in the dataset are licensed under the Creative Commons Attribution 4.0 International License (CC BY 4.0).

For proper attribution when using this dataset in any publications or research outputs, please cite with the DOI. 

\textbf{\textit{Suggested Citation:}} Yang, S., Wang, H., Fu, Y., Tian, Y., Kamishima, T., Ikebe, M., Ou, Y., \& Okutomi, M.(2025). RAM-W600: A Multi-Task Wrist Dataset and
Benchmark for Rheumatoid Arthritis. \url{https://doi.org/10.57967/hf/5328}

\subsection{Data Rights Compliance and Issue Reporting} \label{sec:data_protection}

We are committed to complying data protection rights in accordance with relevant regulations, including but not limited to the General Data Protection Regulation (GDPR). All personally identifiable information (PII) has been removed through anonymization techniques. If any individual represented in the dataset wishes to have their data removed, we provide a clear and accessible process for issue reporting and resolution via our GitHub repository. Concerned parties are encouraged to contact the authors directly through the contact form linked on the GitHub page. Upon receiving a request, we will engage with the individual to verify their identity and promptly remove the relevant data entries from the dataset.

\begin{table*}[!t]
\small
\caption{Radiographic imaging configuration parameters}
\label{tab:madical_info_AP}
\centering
\begin{threeparttable}
\setlength{\tabcolsep}{0mm}{
\begin{tabular}{p{3cm}<{\centering}p{1.8cm}<{\centering}p{1.8cm}<{\centering}p{1.8cm}<{\centering}p{2.8cm}<{\centering}p{1.2cm}<{\centering}p{1.6cm}<{\centering}}
\toprule
& \textbf{HU} & \textbf{HMCRD} & \textbf{SCGH} & \textbf{DHA} & \textbf{BTXRD} & \textbf{FA}  \\
\midrule
Model & - & Radnext 32 & KXO-50G & IPI LAB(Secondary) & - & -\\
Manufacturer & FUJIFILM & HITACHI & TOSHIBA & Array(Secondary) & - & FUJIFILM \& Philips\\
Aluminum filter (mm) & NO & 0.5 & NO & - & - & -\\
Tube voltage (kV) & - & 50 & 45 & -  & - & -\\
Tube current (mA) & - & 100 & 250 & -  & - & -\\
Exposure time (mSec) & - & 25 & 14 & -  & - & -\\
Source to image (cm) & - & 100 & 100 & - & - & -\\
Resolution (mm/pixel) & 0.15 & 0.15 & 0.15 & - & - & -\\
Image size (pixel) & 2010$\times$1670 & 2010$\times$1490 & 2010$\times$1490 & 1744$\times$2126  & - & -\\
Bit depth (bit) & 16 & 10 & 10 & 16  & - & -\\
\bottomrule
\end{tabular}
}
\raggedright
\begin{tablenotes}
\item \textbf{HU}: Faculty of Health Sciences, Hokkaido University.
\item \textbf{HMCRD}: Hokkaido Medical Center for Rheumatic Diseases, Japan.
\item \textbf{SCGH}: Sapporo City General Hospital, Japan.
\item \textbf{DHA}: Digital Hand Atlas, University of Southern California, US.
\item \textbf{BTXRD}: Bone Tumor X-ray Radiograph Dataset, Biomedicine Discovery Institute and Department of Biochemistry and Molecular Biology, Monash University, Australia. 
\item \textbf{FA}: FracAtlas, Islamic University of Technology, Bangladesh. 
\end{tablenotes}
\end{threeparttable}
\end{table*}

\subsection{Data Acquisition}
Radiographs were collected from six institutions with varying imaging configurations, including differences in equipment models, acquisition settings, and image resolutions, as shown in Table~\ref{tab:madical_info_AP}.

\begin{figure}[!t]
    \centering
    \includegraphics[width=1\linewidth]{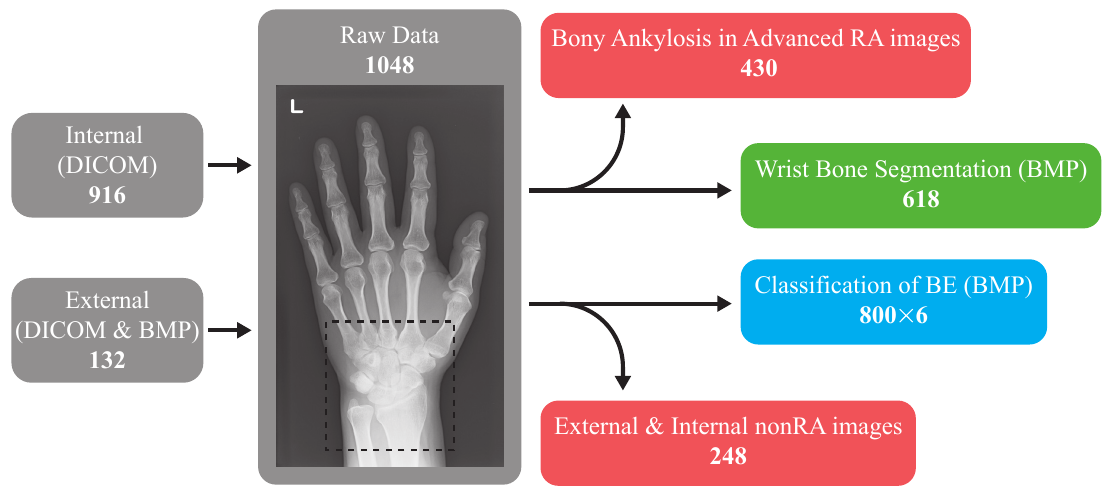}
    \caption{A total of 1048 DICOM-format wrist radiographs were collected, including 916 internal cases from our institutions and 132 external cases from three different sources. Within the internal cohort, 116 images were identified as non-RA, while the remaining were RA cases. All 132 external images were non-RA. After filtering, 430 advanced RA cases with bony ankylosis were excluded. The final dataset was used for two primary tasks: wrist bone instance segmentation (618 BMP images) and BE classification (800 images × 6 joint areas). The external non-RA images were used exclusively for comparison purposes.}
    \label{fig:data_filter}
\end{figure}

\subsection{Data Pre-Processing}
\label{sec:data_pre}

\begin{table*}[!t]
\caption{Joint score distribution across train, valid, and test sets of RAM-W600.}
\centering
\begin{subtable}[t]{0.45\textwidth}
\centering
\caption{\textbf{\textit{Wrist Bone Segmentation}} Dataset}
\resizebox{\linewidth}{!}{
\begin{tabular}{ccccccc}
\toprule
\textbf{Score} & \textbf{MC 1} & \textbf{Tr} & \textbf{Sca} & \textbf{Lu} & \textbf{Radius} & \textbf{Ulna} \\
\midrule
\multicolumn{7}{c}{\textit{Train Set}} \\
\midrule
0 & 414 & 384 & 373 & 421 & 423 & 400 \\
1 & 5  & 8  & 14  & 2  & 0  & 9  \\
2 & 6  & 30  & 34  & 2  & 2  & 15  \\
3 & 0  & 3  & 4  & 0  & 0  & 1  \\
5 & 0  & 0  & 0  & 0  & 0  & 0  \\
\midrule
\multicolumn{7}{c}{\textit{Valid Set}} \\
\midrule
0 & 66 & 61 & 56 & 69 & 66 & 59 \\
1 & 1  & 3  & 6  & 0  & 3  & 0  \\
2 & 2  & 5 & 7 & 0  & 0  & 10  \\
3 & 0  & 0  & 0  & 0  & 0  & 0  \\
5 & 0  & 0  & 0  & 0  & 0  & 0  \\
\midrule
\multicolumn{7}{c}{\textit{Test Set}} \\
\midrule
0 & 117 & 110 & 110 & 121 & 120 & 116 \\
1 & 4  & 4  & 0  & 1  & 1  & 2  \\
2 & 3  & 8 & 11 & 2  & 3  & 6   \\
3 & 0  & 2  & 3  & 0  & 0  & 0  \\
5 & 0  & 0  & 0  & 0  & 0  & 0  \\
\bottomrule
\end{tabular}}
\end{subtable}
\hspace{0.2cm} 
\begin{subtable}[t]{0.4545\textwidth}
\centering
\caption{\textbf{\textit{SvdH BE Scoring}} Dataset}
\resizebox{\linewidth}{!}{
\begin{tabular}{ccccccc}
\toprule
\textbf{Score} & \textbf{MC 1} & \textbf{Tr} & \textbf{Sca} & \textbf{Lu} & \textbf{Radius} & \textbf{Ulna} \\
\midrule
\multicolumn{7}{c}{\textit{Train Set}} \\
\midrule
0 & 493 & 357 & 338 & 531 & 541 & 470 \\
1 & 55 & 117 & 123 & 11 & 8 & 37 \\
2 & 11 & 83 & 79 & 8 & 4 & 38 \\
3 & 0 & 2 & 16 & 5 & 6 & 0 \\
5 & 0 & 0 & 3 & 4 & 0 & 4 \\
\midrule
\multicolumn{7}{c}{\textit{Valid Set}} \\
\midrule
0 & 67 & 42 & 49 & 74 & 68 & 58 \\
1 & 10 & 20 & 11 & 2 & 6 & 11 \\
2 & 4 & 18 & 16 & 1 & 3 & 8 \\
3 & 0 & 1 & 3 & 0 & 4 & 3 \\
5 & 0 & 0 & 2 & 4 & 0 & 1 \\
\midrule
\multicolumn{7}{c}{\textit{Test Set}} \\
\midrule
0 & 140 & 90 & 94 & 157 & 151 & 123 \\
1 & 18 & 39 & 36 & 2 & 7 & 21 \\
2 & 2 & 28 & 26 & 0 & 1 & 13 \\
3 & 0 & 3 & 4 & 0 & 1 & 3 \\
5 & 0 & 0 & 0 & 1 & 0 & 0 \\
\bottomrule
\end{tabular}
}
\end{subtable}
\label{tab:dataset_distribution}
\end{table*}

\begin{table*}[!t]
\caption{Institution score distribution across train, valid, and test sets of RAM-W600.}
\centering
\begin{subtable}[t]{0.58\textwidth}
\centering
\caption{\textbf{\textit{Wrist Bone Segmentation}} Dataset}
\label{tab:dataset_distribution_institution_seg}
\resizebox{\linewidth}{!}{
\begin{tabular}{ccccccc}
\toprule
\textbf{Score} & \textbf{HMCRD} & \textbf{SCGH} & \textbf{HU} & \textbf{DHA} & \textbf{BTXRD} & \textbf{FA} \\
\midrule
\multicolumn{7}{c}{\textit{Train Set}} \\
\midrule
NonRA & 540 & 0 & 24 & 318 & 210 & 36 \\
0 & 1038 & 208 & 41 & 0 & 0 & 0 \\
1 & 26  & 9  & 3  & 0  & 0  & 0  \\
$\geq$2 & 76  & 17  & 4  & 0  & 0  & 0  \\
\midrule
\multicolumn{7}{c}{\textit{Valid Set}} \\
\midrule
NonRA & 24 & 0 & 12 & 48 & 24 & 12 \\
0 & 212 & 21 & 24 & 0 & 0 & 0 \\
1 & 13  & 0  & 0  & 0  & 0  & 0  \\
$\geq$2 & 21  & 3 & 0 & 0  & 0  & 0  \\

\midrule
\multicolumn{7}{c}{\textit{Test Set}} \\
\midrule
NonRA & 96 & 0 & 0 & 72 & 48 & 24 \\
0 & 354 & 100 & 0 & 0 & 0 & 0 \\
1 & 8  & 4  & 0  & 0  & 0  & 0  \\
$\geq$2 & 28  & 10 & 0 & 0  & 0  & 0   \\

\bottomrule
\end{tabular}}
\end{subtable}
\hspace{0.2cm} 
\begin{subtable}[t]{0.3515\textwidth}
\centering
\caption{\textbf{\textit{SvdH BE Scoring}} Dataset}
\label{tab:dataset_distribution_institution_cls}
\resizebox{\linewidth}{!}{
\begin{tabular}{cccc}
\toprule
\textbf{Score} & \textbf{HMCRD} & \textbf{SCGH} & \textbf{HU} \\
\midrule
\multicolumn{4}{c}{\textit{Train Set}} \\
\midrule
0 & 1628 & 353 & 744 \\
1 & 93 & 32 & 226 \\
$\geq$2 & 103 & 41 & 128 \\
\midrule
\multicolumn{4}{c}{\textit{Valid Set}} \\
\midrule
0 & 256 & 18 & 84 \\
1 & 15 & 5 & 40 \\
$\geq$2 & 23 & 1 & 44 \\
\midrule
\multicolumn{4}{c}{\textit{Test Set}} \\
\midrule
0 & 408 & 92 & 255 \\
1 & 21 & 8 & 94 \\
$\geq$2 & 39 & 8 & 35 \\
\bottomrule
\end{tabular}
}
\end{subtable}
\label{tab:dataset_distribution_institution}
\end{table*}

In the pre-processing pipeline of RAM-W600 (Fig.~\ref{fig:data_filter}), we first localize the ROI around the wrist across all 1048 DICOM-format hand radiographs. For the wrist bone segmentation task, we exclude 430 images exhibiting bony ankylosis associated with advanced RA, resulting in a curated subset of 618 BMP-format images for segmentation. For the BE classification task, we exclude only 248 non-RA images from the internal and external cohorts, retaining 800 RA cases from our internal dataset. These cases are subsequently converted to BMP format, and six joint-level crops are extracted per image, yielding a total of 4800 samples for BE classification.


The wrist bone segmentation dataset and the SvdH BE scoring dataset are split independently. To prevent data leakage and reduce potential bias, we randomly partition the cases into training, validation, and test sets based on unique patient IDs using an approximate ratio of 70\%/10\%/20\%. Table~\ref{tab:dataset_distribution} summarizes the distribution of the six wrist joints (1st Metacarpal, Trapezium, Scaphoid, Lunate, Radius, and Ulna) across scores 0–5 within the training, validation, and test subsets of both datasets. The majority of joints are assigned an SvdH BE score of 0, while those with a score of 5 are extremely rare. In both tasks, the number of joints decreases as the score increases, resulting in a clearly imbalanced distribution. 

In addition, Table~\ref{tab:dataset_distribution_institution} further details the institution-wise distribution of cases across the two datasets. Table~\ref{tab:dataset_distribution_institution_seg} presents the distribution for the wrist bone segmentation task, where both RA and nonRA cases are included, while Table~\ref{tab:dataset_distribution_institution_cls} shows the corresponding distribution of joint scores in the SvdH BE scoring dataset. This breakdown highlights the contribution of each collaborating institution and illustrates how score imbalance manifests across different sources and subsets.

\begin{figure}[!t]
    \centering
    \includegraphics[width=1\linewidth]{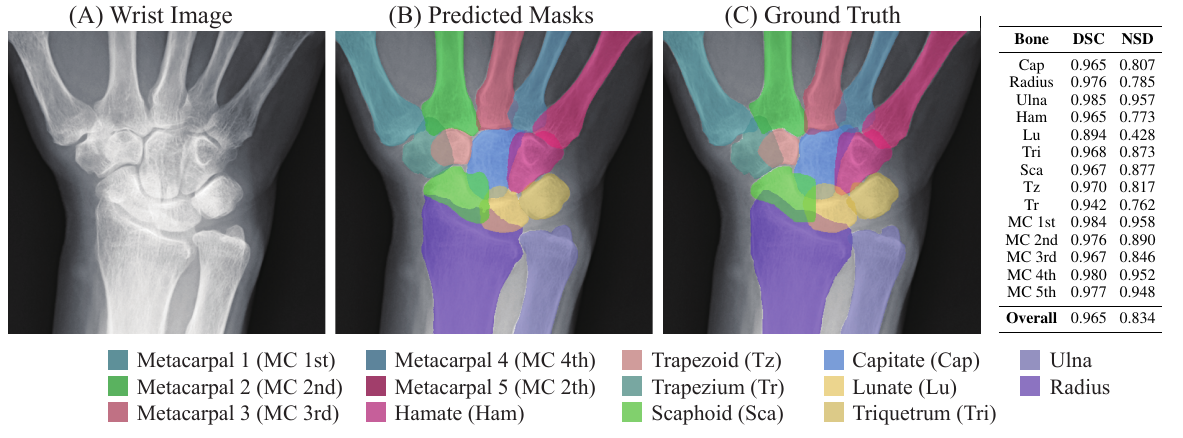}
    \caption{Wrist bone segmentation.(A) Original wrist radiograph. (B) Predicted instance segmentation masks. (C) Ground truth annotations. The right panel reports the segmentation performance per bone.}
    \label{fig:seg_task_intro_AP}
\end{figure}

\subsection{Dataset Maintenance}
As the authors and maintainers of this dataset, we affirm that while the dataset is self-contained and does not depend on any external links or content, we may provide future updates, such as adding new cases or incorporating additional tasks. These potential updates aim to enhance the dataset's value while maintaining its long-term usability.

\begin{figure}[!t]
    \centering
    \includegraphics[width=1\linewidth]{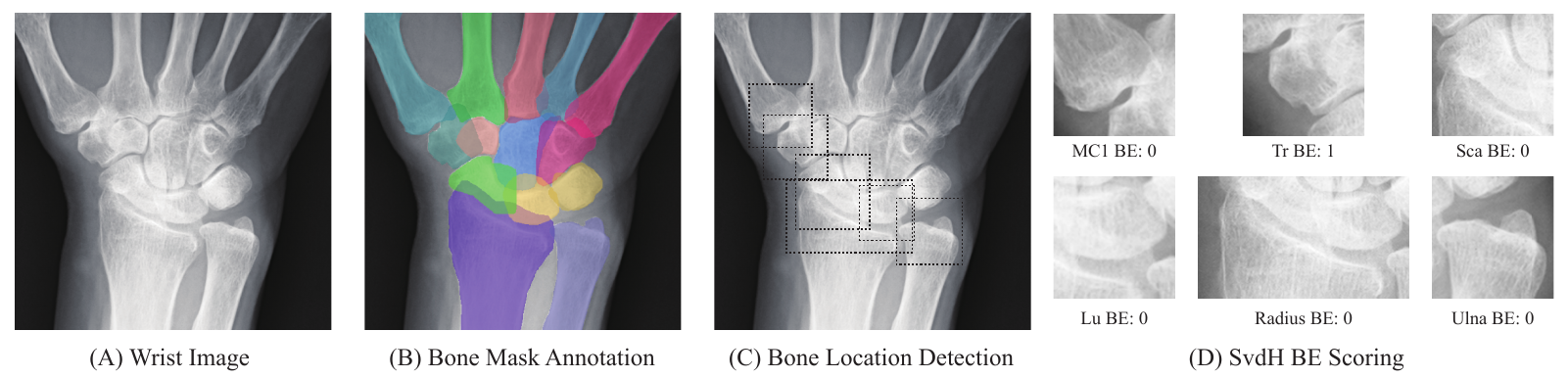}
    \caption{Image input and annotation.(A) Raw wrist radiograph. (B) Instance bone segmentation mask annotation (C) bone location annotations for target regions. (D) SvdH BE scores assigned to each joint region.}
    \label{fig:intro_input}
\end{figure}

\subsection{Wrist Bone Segmentation}
Wrist bone segmentation from radiographs is a critical prerequisite for downstream tasks such as joint localization, morphological analysis, and BE scoring in RA assessment. As illustrated in Fig.~\ref{fig:seg_task_intro_AP}, this task involves delineating multiple overlapping and irregularly shaped carpal and metacarpal bones, which often exhibit low contrast and anatomical ambiguity in radiographs. Accurate segmentation enables reliable quantification of structural features and supports automated interpretation in clinical workflows.

\begin{figure}[!t]
    \centering
    \includegraphics[width=0.9\linewidth]{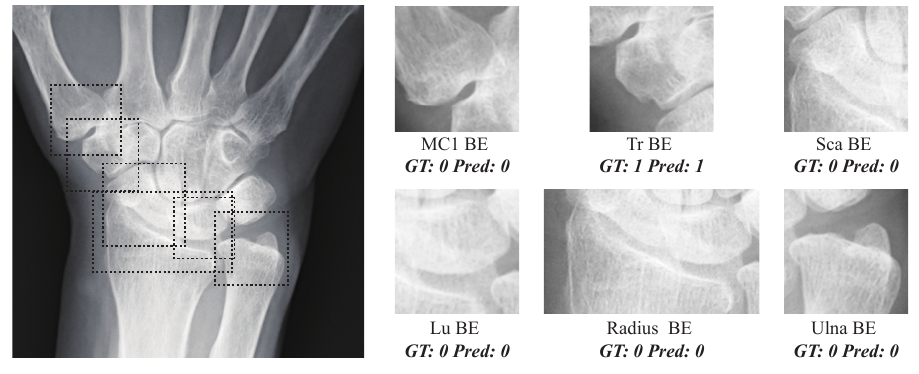}
    \caption{BE and nonBE Classification. The left panel shows six annotated joint regions used for BE classification. The right panels display each joint with ground truth (GT) and predicted (Pred) SvdH BE scores.}
    \label{fig:cls_task_intro_AP}
\end{figure}

In this task, we annotate 14 distinct wrist bones, including both carpal, metacarpal components and Distal Radius $\&$ Distal Ulna. Notably, the Pisiform and Triquetrum bones are difficult to distinguish in clinical practice due to their overlapping appearance and low visibility on standard radiographs. Consequently, it is challenging to evaluate them as independent diagnostic regions~\cite{nanduri2022triquetrum}. Therefore, we merge these two structures into a single category during annotation to reflect their practical indistinguishability. The input to the segmentation model is the wrist ROI cropped from the radiograph, and the output and ground truth are a pixel-wise mask for each annotated bone, as illustrated in Fig.~\ref{fig:intro_input}.

\subsection{Classification of BE}
BE classification is a key component of the SvdH scoring system, widely adopted in clinical practice for evaluating joint damage in RA. As illustrated in Fig.~\ref{fig:cls_task_intro_AP}, this task involves identifying subtle pathological changes in individual carpal bones from radiographs, such as cortical breaks and irregular bone surfaces. The classification task is particularly challenging due to the subtlety of erosion features and the high degree of anatomical overlap in wrist joints. Accurate BE detection is essential for automated RA scoring systems and downstream severity assessment, yet remains difficult for both traditional and deep learning models, especially under class imbalance and in early-stage lesions.

In this task, we annotate the SvdH BE scores for six joint surfaces within the wrist. To formulate the problem as a binary classification task, all joint surfaces with non-zero scores were treated as positive cases (i.e., exhibiting BE), while those with a score of zero were treated as negative cases (i.e., without BE). The input to the model is the ROI corresponding to an individual joint surface, and the output is a probability distribution over the two classes, representing the model’s confidence in the presence or absence of BE, as shown in Fig.~\ref{fig:intro_input}.


\begin{table*}[!t]
\centering
\caption{DSC performance on all joints. 
The best results in each column are highlighted in \textbf{bold}, 
and the second-best values are \underline{underlined}.}
\label{tab:mean_joint_dsc}

\begin{subtable}{\textwidth}
\centering
\resizebox{\textwidth}{!}{
\begin{tabular}{lcccccccc}
\toprule
\textbf{Model} & \textbf{Cap} & \textbf{Radius} & \textbf{Ulna} & \textbf{Ham} & \textbf{Lu} & \textbf{Tri} & \textbf{Sca} \\
\midrule
\multicolumn{8}{c}{\textbf{Supervised Models}}  \\
\midrule
UNet        & 96.24$\pm$0.11 & 98.17$\pm$0.05 & 98.20$\pm$0.06 & 96.25$\pm$0.09 & 95.29$\pm$0.19 & 96.37$\pm$0.08 & 96.80$\pm$0.13 \\
DeepLabV3   & 96.49$\pm$0.08 & 98.32$\pm$0.04 & 97.98$\pm$0.02 & 96.26$\pm$0.04 & 95.47$\pm$0.06 & 96.20$\pm$0.06 & 96.68$\pm$0.06 \\
FPN         & 96.32$\pm$0.08 & 98.21$\pm$0.03 & 98.22$\pm$0.06 & 96.08$\pm$0.09 & 95.43$\pm$0.12 & 96.26$\pm$0.13 & 96.45$\pm$0.12 \\
PSPNet      & 94.91$\pm$0.13 & 97.39$\pm$0.15 & 96.87$\pm$0.20 & 94.78$\pm$0.05 & 93.51$\pm$0.20 & 94.93$\pm$0.09 & 94.69$\pm$0.26 \\
DeepLabV3+  & 96.50$\pm$0.08 & 98.39$\pm$0.01 & 98.44$\pm$0.03 & 96.21$\pm$0.06 & 95.52$\pm$0.03 & 96.55$\pm$0.04 & 96.73$\pm$0.07 \\
SegResNet   & 95.68$\pm$0.31 & 97.97$\pm$0.15 & 98.16$\pm$0.03 & 95.84$\pm$0.22 & 95.10$\pm$0.25 & 96.00$\pm$0.20 & 96.60$\pm$0.24 \\
UNet++      & \underline{97.06$\pm$0.08} & 98.46$\pm$0.06 & 98.55$\pm$0.13 & 96.75$\pm$0.04 & 95.91$\pm$0.13 & 97.11$\pm$0.09 & 97.28$\pm$0.05 \\
SegFormer   & 96.63$\pm$0.06 & 98.34$\pm$0.04 & 98.37$\pm$0.04 & 96.35$\pm$0.08 & 95.78$\pm$0.09 & 96.56$\pm$0.08 & 96.84$\pm$0.08 \\
TransUNet   & 97.41$\pm$0.10 & \underline{98.71$\pm$0.07} & \underline{98.88$\pm$0.02} & \underline{97.18$\pm$0.03} & \underline{96.51$\pm$0.06} & 97.30$\pm$0.06 & \underline{97.73$\pm$0.02} \\
UKAN        & 96.32$\pm$0.16 & 97.97$\pm$0.73 & 98.44$\pm$0.06 & 96.26$\pm$0.16 & 95.64$\pm$0.09 & 96.36$\pm$0.10 & 96.72$\pm$0.08 \\
UMambaBot   & 97.32$\pm$0.04 & 98.55$\pm$0.05 & 98.75$\pm$0.02 & 97.04$\pm$0.04 & 96.26$\pm$0.07 & 97.23$\pm$0.09 & 97.59$\pm$0.03 \\
UMambaEnc   & 97.33$\pm$0.06 & 98.56$\pm$0.12 & 98.81$\pm$0.03 & 97.00$\pm$0.05 & 96.38$\pm$0.10 & \underline{97.40$\pm$0.08} & 97.59$\pm$0.10 \\
SwinUMamba  & \textbf{97.58$\pm$0.02} & \textbf{98.84$\pm$0.02} & \textbf{98.91$\pm$0.06} & \textbf{97.29$\pm$0.05} & \textbf{96.66$\pm$0.09} & \textbf{97.50$\pm$0.06} & \textbf{97.85$\pm$0.03} \\
\midrule
\multicolumn{8}{c}{\textbf{Foundation Models}}  \\
\midrule
SAM (box)     & 90.67$\pm$3.68 & 92.49$\pm$2.35 & 93.21$\pm$11.63 & 87.29$\pm$3.57 & 83.38$\pm$5.17 & 92.72$\pm$3.71 & 87.64$\pm$5.09 \\
SAM (pt)      & 75.64$\pm$27.72 & 92.04$\pm$6.54 & 96.56$\pm$5.28 & 73.56$\pm$22.31 & 81.30$\pm$13.21 & 91.08$\pm$8.03 & 79.09$\pm$18.78 \\
MedSAM (box)  & 82.54$\pm$6.60 & 90.63$\pm$2.88 & 94.91$\pm$4.64 & 82.43$\pm$5.70 & 80.06$\pm$5.91 & 88.18$\pm$4.11 & 82.05$\pm$6.47 \\
\bottomrule
\end{tabular}}
\end{subtable}

\vspace{0.6cm}

\begin{subtable}{\textwidth}
\centering
\resizebox{\textwidth}{!}{
\begin{tabular}{lcccccccc}
\toprule
\textbf{Model} & \textbf{Tr} & \textbf{Tz} & \textbf{MC1} & \textbf{MC2} & \textbf{MC3} & \textbf{MC4} & \textbf{MC5} \\
\midrule
\multicolumn{8}{c}{\textbf{Supervised Models}}  \\
\midrule
UNet        & 95.69$\pm$0.19 & 94.14$\pm$0.08 & 98.14$\pm$0.09 & 97.72$\pm$0.27 & 96.87$\pm$0.09 & 97.30$\pm$0.07 & 97.90$\pm$0.07 \\
DeepLabV3   & 95.53$\pm$0.08 & 93.98$\pm$0.07 & 97.98$\pm$0.02 & 97.77$\pm$0.08 & 97.00$\pm$0.03 & 97.30$\pm$0.02 & 97.89$\pm$0.03 \\
FPN         & 95.43$\pm$0.11 & 93.86$\pm$0.07 & 97.90$\pm$0.10 & 97.96$\pm$0.05 & 97.14$\pm$0.09 & 97.53$\pm$0.06 & 98.06$\pm$0.05 \\
PSPNet      & 94.12$\pm$0.09 & 92.90$\pm$0.08 & 96.67$\pm$0.08 & 96.94$\pm$0.02 & 95.90$\pm$0.12 & 96.17$\pm$0.05 & 96.96$\pm$0.14 \\
DeepLabV3+  & 95.67$\pm$0.06 & 94.04$\pm$0.07 & 98.21$\pm$0.02 & 98.01$\pm$0.04 & 97.18$\pm$0.02 & 97.62$\pm$0.03 & 98.18$\pm$0.02 \\
SegResNet   & 95.36$\pm$0.30 & 94.07$\pm$0.13 & 97.93$\pm$0.25 & 97.22$\pm$1.10 & 96.86$\pm$0.13 & 97.48$\pm$0.17 & 98.05$\pm$0.09 \\
UNet++      & 96.16$\pm$0.08 & 94.62$\pm$0.07 & 98.39$\pm$0.09 & 98.32$\pm$0.07 & 97.52$\pm$0.04 & 98.03$\pm$0.03 & 98.44$\pm$0.05 \\
SegFormer   & 95.75$\pm$0.06 & 94.24$\pm$0.10 & 98.16$\pm$0.03 & 98.07$\pm$0.04 & 97.20$\pm$0.04 & 97.72$\pm$0.02 & 98.17$\pm$0.03 \\
TransUNet   & 96.54$\pm$0.04 & 95.05$\pm$0.09 & \underline{98.64$\pm$0.06} & \underline{98.45$\pm$0.15} & 97.65$\pm$0.08 & \underline{98.16$\pm$0.16} & \underline{98.49$\pm$0.21} \\
UKAN        & 95.70$\pm$0.05 & 94.20$\pm$0.07 & 98.16$\pm$0.07 & 97.96$\pm$0.08 & 97.16$\pm$0.06 & 97.69$\pm$0.05 & 98.19$\pm$0.05 \\
UMambaBot   & 96.42$\pm$0.07 & \underline{94.85$\pm$0.05} & 98.56$\pm$0.04 & 98.42$\pm$0.03 & \underline{97.69$\pm$0.04} & 98.21$\pm$0.02 & 98.55$\pm$0.02 \\
UMambaEnc   & \underline{96.50$\pm$0.06} & 94.90$\pm$0.11 & 98.60$\pm$0.08 & 98.44$\pm$0.04 & \underline{97.69$\pm$0.05} & 98.13$\pm$0.08 & 98.54$\pm$0.06 \\
SwinUMamba  & \textbf{96.67$\pm$0.02} & \textbf{95.14$\pm$0.04} & \textbf{98.74$\pm$0.04} & \textbf{98.57$\pm$0.01} & \textbf{97.84$\pm$0.02} & \textbf{98.31$\pm$0.01} & \textbf{98.64$\pm$0.01} \\
\midrule
\multicolumn{8}{c}{\textbf{Foundation Models}}  \\
\midrule
SAM (box)     & 83.89$\pm$7.05 & 85.81$\pm$6.99 & 97.13$\pm$1.06 & 91.08$\pm$18.93 & 88.96$\pm$14.95 & 78.39$\pm$31.39 & 89.68$\pm$20.65 \\
SAM (pt)      & 66.05$\pm$22.68 & 59.37$\pm$19.22 & 96.90$\pm$1.74 & 83.90$\pm$20.71 & 86.35$\pm$17.22 & 76.50$\pm$22.18 & 67.02$\pm$25.35 \\
MedSAM (box)  & 82.43$\pm$5.49 & 74.59$\pm$9.48 & 92.76$\pm$6.80 & 87.06$\pm$6.66 & 85.95$\pm$4.00 & 80.18$\pm$8.83 & 87.14$\pm$6.02 \\
\bottomrule
\end{tabular}}
\begin{tablenotes}
\item Foundation models: one inference (mean $\pm$ std across cases).\newline
Supervised models: five runs (mean $\pm$ std across runs).
\end{tablenotes}
\end{subtable}

\end{table*}

\begin{table*}[!t]
\centering
\caption{NSD performance on all joints. The best results in each column are highlighted in \textbf{bold}, and the second-best values are \underline{underlined}.}
\label{tab:mean_joint_nsd}

\begin{subtable}{\textwidth}
\centering
\resizebox{\textwidth}{!}{
\begin{tabular}{lcccccccc}
\toprule
\textbf{Model} & \textbf{Cap} & \textbf{Radius} & \textbf{Ulna} & \textbf{Ham} & \textbf{Lu} & \textbf{Tri} & \textbf{Sca} \\
\midrule
\multicolumn{8}{c}{\textbf{Supervised Models}}  \\
\midrule
UNet        & 75.30$\pm$1.04 & 82.16$\pm$0.43 & 94.05$\pm$0.44 & 78.34$\pm$0.66 & 74.95$\pm$0.64 & 82.69$\pm$0.37 & 83.13$\pm$1.17 \\
DeepLabV3   & 74.77$\pm$0.94 & 82.18$\pm$0.78 & 93.29$\pm$0.15 & 76.32$\pm$0.66 & 74.40$\pm$0.56 & 80.29$\pm$0.52 & 80.74$\pm$0.53 \\
FPN         & 73.08$\pm$0.88 & 81.14$\pm$0.47 & 93.94$\pm$0.39 & 74.62$\pm$1.01 & 73.40$\pm$0.94 & 80.32$\pm$1.44 & 78.43$\pm$1.33 \\
PSPNet      & 59.43$\pm$1.20 & 70.27$\pm$1.72 & 83.14$\pm$2.08 & 62.68$\pm$0.50 & 61.55$\pm$0.94 & 68.70$\pm$1.09 & 63.22$\pm$1.85 \\
DeepLabV3+  & 75.29$\pm$0.94 & 83.44$\pm$0.32 & 96.17$\pm$0.28 & 76.28$\pm$0.58 & 74.05$\pm$0.35 & 83.56$\pm$0.38 & 81.78$\pm$0.89 \\
SegResNet   & 70.71$\pm$2.29 & 80.15$\pm$1.62 & 93.92$\pm$0.35 & 75.30$\pm$1.38 & 73.49$\pm$1.73 & 79.72$\pm$1.86 & 81.10$\pm$1.95 \\
UNet++      & \underline{82.04$\pm$0.82} & 85.22$\pm$1.26 & 96.05$\pm$0.41 & 81.97$\pm$0.35 & 78.28$\pm$0.86 & 88.78$\pm$0.36 & 87.00$\pm$0.60 \\
SegFormer   & 76.97$\pm$0.83 & 82.99$\pm$0.71 & 95.77$\pm$0.27 & 78.06$\pm$1.01 & 76.49$\pm$0.81 & 84.17$\pm$0.55 & 82.60$\pm$0.96 \\
TransUNet   & \underline{84.82$\pm$0.94} & \underline{89.16$\pm$0.70} & \underline{98.16$\pm$0.15} & \underline{85.98$\pm$0.43} & \underline{82.21$\pm$0.69} & 90.12$\pm$0.33 & \underline{90.96$\pm$0.33} \\
UKAN        & 74.13$\pm$1.17 & 81.26$\pm$2.53 & 95.25$\pm$0.41 & 77.15$\pm$1.03 & 75.89$\pm$0.87 & 82.19$\pm$0.82 & 81.38$\pm$0.79 \\
UMambaBot   & 84.19$\pm$0.41 & 87.19$\pm$0.94 & 97.59$\pm$0.12 & 84.93$\pm$0.24 & 81.00$\pm$0.48 & 90.22$\pm$0.73 & 89.99$\pm$0.14 \\
UMambaEnc   & 84.17$\pm$0.68 & 87.62$\pm$1.47 & 97.80$\pm$0.20 & 84.55$\pm$0.68 & 81.52$\pm$0.70 & \underline{91.02$\pm$0.45} & 90.06$\pm$0.55 \\
SwinUMamba  & \textbf{86.71$\pm$0.20} & \textbf{90.86$\pm$0.33} & \textbf{98.24$\pm$0.30} & \textbf{87.31$\pm$0.47} & \textbf{83.75$\pm$0.58} & \textbf{92.29$\pm$0.37} & \textbf{91.99$\pm$0.17} \\
\midrule
\multicolumn{8}{c}{\textbf{Foundation Models}}  \\
\midrule
SAM (box)     & 57.24$\pm$14.60 & 61.51$\pm$8.00 & 82.09$\pm$14.87 & 46.82$\pm$11.76 & 56.04$\pm$11.46 & 65.82$\pm$16.59 & 63.67$\pm$10.05 \\
SAM (pt)      & 46.48$\pm$28.63 & 62.62$\pm$12.69 & 91.68$\pm$12.18 & 32.01$\pm$20.09 & 54.22$\pm$16.56 & 66.38$\pm$21.09 & 53.52$\pm$22.49 \\
MedSAM (box)  & 20.81$\pm$13.22 & 45.31$\pm$13.06 & 75.01$\pm$19.14 & 18.82$\pm$10.95 & 38.20$\pm$17.00 & 37.38$\pm$16.40 & 26.78$\pm$12.72 \\
\bottomrule
\end{tabular}}
\end{subtable}

\vspace{0.6cm}

\begin{subtable}{\textwidth}
\centering
\resizebox{\textwidth}{!}{
\begin{tabular}{lcccccccc}
\toprule
\textbf{Model} & \textbf{Tr} & \textbf{Tz} & \textbf{MC1} & \textbf{MC2} & \textbf{MC3} & \textbf{MC4} & \textbf{MC5} \\
\midrule
\multicolumn{8}{c}{\textbf{Supervised Models}}  \\
\midrule
UNet        & 74.47$\pm$1.52 & 70.46$\pm$0.69 & 92.91$\pm$0.59 & 90.64$\pm$1.62 & 84.13$\pm$0.64 & 90.07$\pm$0.50 & 93.75$\pm$0.15 \\
DeepLabV3   & 72.75$\pm$0.80 & 68.31$\pm$0.55 & 92.25$\pm$0.11 & 91.87$\pm$0.38 & 85.46$\pm$0.14 & 91.30$\pm$0.19 & 93.75$\pm$0.29 \\
FPN         & 71.15$\pm$0.97 & 66.85$\pm$1.08 & 90.36$\pm$0.78 & 91.89$\pm$0.38 & 84.73$\pm$0.44 & 90.79$\pm$0.44 & 93.48$\pm$0.59 \\
PSPNet      & 59.64$\pm$1.22 & 59.15$\pm$0.47 & 81.49$\pm$0.62 & 84.69$\pm$0.26 & 75.78$\pm$1.35 & 81.37$\pm$0.44 & 85.32$\pm$0.62 \\
DeepLabV3+  & 73.86$\pm$0.61 & 68.78$\pm$0.70 & 93.53$\pm$0.27 & 92.48$\pm$0.23 & 85.45$\pm$0.19 & 91.69$\pm$0.14 & 95.20$\pm$0.18 \\
SegResNet   & 71.99$\pm$2.26 & 70.53$\pm$1.33 & 91.49$\pm$1.42 & 88.49$\pm$3.81 & 83.46$\pm$0.67 & 90.48$\pm$1.14 & 94.20$\pm$0.58 \\
UNet++      & 77.99$\pm$0.81 & 74.07$\pm$0.49 & 94.44$\pm$0.70 & 94.19$\pm$0.48 & 87.59$\pm$0.21 & 93.95$\pm$0.30 & 96.22$\pm$0.27 \\
SegFormer   & 74.57$\pm$0.54 & 70.75$\pm$0.86 & 93.40$\pm$0.36 & 93.31$\pm$0.15 & 86.16$\pm$0.36 & 92.89$\pm$0.13 & 95.31$\pm$0.21 \\
TransUNet   & 81.55$\pm$0.30 & 77.19$\pm$0.81 & \underline{96.20$\pm$0.85} & \underline{95.62$\pm$0.35} & 88.73$\pm$0.39 & \underline{95.77$\pm$0.10} & 96.27$\pm$3.00 \\
UKAN        & 73.97$\pm$0.65 & 69.99$\pm$0.92 & 92.62$\pm$0.35 & 91.38$\pm$0.59 & 85.07$\pm$0.41 & 91.83$\pm$0.36 & 94.59$\pm$0.44 \\
UMambaBot   & 81.20$\pm$0.67 & \underline{75.52$\pm$0.66} & 95.95$\pm$0.22 & 94.99$\pm$0.32 & \underline{88.83$\pm$0.21} & 95.63$\pm$0.10 & \underline{97.22$\pm$0.08} \\
UMambaEnc   & \underline{82.07$\pm$0.50} & 76.15$\pm$1.04 & 96.34$\pm$0.33 & 95.13$\pm$0.38 & 88.73$\pm$0.38 & 95.07$\pm$0.62 & 97.14$\pm$0.25 \\
SwinUMamba  & \textbf{83.21$\pm$0.13} & \textbf{77.95$\pm$0.61} & \textbf{97.34$\pm$0.28} & \textbf{96.15$\pm$0.15} & \textbf{89.92$\pm$0.16} & \textbf{96.32$\pm$0.06} & \textbf{97.91$\pm$0.06} \\
\midrule
\multicolumn{8}{c}{\textbf{Foundation Models}}  \\
\midrule
SAM (box)     & 47.55$\pm$12.22 & 43.56$\pm$16.30 & 88.81$\pm$5.26 & 78.79$\pm$15.78 & 65.68$\pm$11.48 & 65.60$\pm$20.89 & 78.44$\pm$20.84 \\
SAM (pt)      & 34.34$\pm$20.19 & 23.77$\pm$17.87 & 88.52$\pm$6.65 & 66.97$\pm$23.86 & 60.10$\pm$17.24 & 54.50$\pm$23.73 & 45.50$\pm$27.15 \\
MedSAM (box)  & 32.61$\pm$10.31 & 15.90$\pm$12.79 & 61.62$\pm$12.88 & 48.02$\pm$17.31 & 40.55$\pm$15.58 & 38.63$\pm$14.70 & 43.69$\pm$13.67 \\
\bottomrule
\end{tabular}}
\begin{tablenotes}
\item Foundation models: one inference (mean $\pm$ std across cases).\newline
Supervised models: five runs (mean $\pm$ std across runs).
\end{tablenotes}
\end{subtable}

\end{table*}

\section{Detailed Analysis of Experimental Results}

\subsection{Wrist Bone Segmentation}
\subsubsection{Overall Segmentation Results}
Table~\ref{tab:mean_joint_dsc} and Table~\ref{tab:mean_joint_nsd} report the segmentation results in terms of DSC and NSD, which together provide complementary perspectives on overlap accuracy and boundary precision. A consistent observation across both metrics is that supervised models substantially outperform foundation models, with Mamba-based architectures leading the performance on nearly all joints. This gap highlights that current foundation models are not sufficient for wrist bone segmentation, making supervised baselines essential for establishing reliable performance benchmarks.

In terms of DSC (Table~\ref{tab:mean_joint_dsc}), supervised models achieve very high overlap accuracy, often above 97\% across most bones. SwinUMamba yields the strongest results, surpassing all other methods in almost every region. For example, it achieves 98.84\% on the Radius, 98.74\% on the MC1, and 98.64\% on the Metacarpal 5th. UMambaEnc and UMambaBot also deliver competitive results, particularly on large and structurally less ambiguous bones such as the Distal Ulna (98.81\% and 98.75\%). Transformer-based models including TransUNet and SegFormer maintain stable performance but are slightly behind the Mamba-based architectures, while CNN-based methods such as UNet and DeepLabV3 show moderate accuracy with greater fluctuations across regions. In sharp contrast, foundation models lag considerably, with SAM (pt) and MedSAM recording DSC values 10 to 20 points lower than supervised models; for instance, SAM (pt) achieves only 75.64\% on the Capitate and 76.50\% on the Metacarpal 4th.

Turning to NSD (Table~\ref{tab:mean_joint_nsd}), which emphasizes surface-level boundary precision, the same trend persists but the performance gap becomes even more pronounced. SwinUMamba again dominates with top results across nearly all joints, showing notable advantages on small and challenging structures such as the Trapezium and Triquetrum where precise boundary delineation is critical. UMambaEnc and UMambaBot remain highly competitive, confirming their robustness across both large bones and finer anatomical details. By comparison, CNN-based models exhibit larger drops in NSD despite acceptable DSC values, indicating difficulties in capturing fine boundary details for complex joint shapes. Foundation models perform the worst under this metric, with SAM- and MedSAM-based approaches consistently trailing far behind, underscoring their limited ability to recover accurate anatomical boundaries in wrist bone segmentation. These findings indicate that overlap-based DSC alone may mask boundary errors, and NSD is necessary to reveal clinically critical differences in fine anatomical structures.

\clearpage

\subsubsection{Impact of BE}

\begin{table*}[!t]
\centering
\caption{DSC performance on representative wrist bones. (Mann-Whitney U test between BE $\&$ nonBE, *: P $<$ 0.05; **: P $<$ 0.01; ***: P $<$ 0.001).}
\label{tab:joint_be_nonbe_dsc}

\begin{subtable}{\textwidth}
\centering
\resizebox{\textwidth}{!}{
\begin{tabular}{l *{3}{ccc}}
\toprule
\multirow{2.5}{*}{\textbf{Model}} & 
\multicolumn{3}{c}{\textbf{Radius}} & 
\multicolumn{3}{c}{\textbf{Ulna}} & 
\multicolumn{3}{c}{\textbf{Lunate}} \\
\cmidrule(lr){2-4}
\cmidrule(lr){5-7}
\cmidrule(lr){8-10}
 & \textbf{BE} & \textbf{nonBE} & \textbf{P} 
 & \textbf{BE} & \textbf{nonBE} & \textbf{P} 
 & \textbf{BE} & \textbf{nonBE} & \textbf{P} \\
\midrule
\multicolumn{10}{c}{\textbf{Supervised Models}}  \\
\midrule
UNet        & 97.96$\pm$0.06 & 98.25$\pm$0.05 & **  & 98.31$\pm$0.03 & 98.16$\pm$0.08 &      & 94.76$\pm$0.08 & 95.49$\pm$0.23 & ***  \\
DeepLabV3   & 98.13$\pm$0.07 & 98.40$\pm$0.04 & *** & 97.92$\pm$0.02 & 98.00$\pm$0.02 &      & 94.61$\pm$0.15 & 95.81$\pm$0.05 & ***  \\
FPN         & 98.04$\pm$0.06 & 98.28$\pm$0.02 & *** & 98.24$\pm$0.06 & 98.21$\pm$0.06 &      & 94.77$\pm$0.14 & 95.69$\pm$0.13 & ***  \\
PSPNet      & 97.28$\pm$0.13 & 97.44$\pm$0.17 & *   & 96.93$\pm$0.23 & 96.85$\pm$0.20 &      & 93.30$\pm$0.17 & 93.59$\pm$0.25 & **   \\
DeepLabV3+  & 98.21$\pm$0.03 & 98.46$\pm$0.02 & *** & 98.45$\pm$0.03 & 98.44$\pm$0.03 &      & 94.85$\pm$0.09 & 95.79$\pm$0.05 & ***  \\
SegResNet   & 97.79$\pm$0.12 & 98.05$\pm$0.16 & **  & 98.24$\pm$0.12 & 98.14$\pm$0.08 &      & 94.73$\pm$0.22 & 95.25$\pm$0.26 & ***  \\
UNet++      & 98.39$\pm$0.05 & 98.48$\pm$0.07 &     & 98.68$\pm$0.10 & 98.49$\pm$0.15 &      & 95.18$\pm$0.11 & 96.20$\pm$0.14 & ***  \\
SegFormer   & 98.17$\pm$0.05 & 98.40$\pm$0.04 & *** & 98.35$\pm$0.04 & 98.37$\pm$0.05 &      & 95.07$\pm$0.12 & 96.06$\pm$0.08 & ***  \\
TransUNet   & 98.63$\pm$0.06 & 98.75$\pm$0.07 & *** & 98.89$\pm$0.02 & 98.87$\pm$0.02 &      & 95.82$\pm$0.09 & 96.78$\pm$0.05 & ***  \\
UKAN        & 97.79$\pm$0.80 & 98.04$\pm$0.69 & **  & 98.44$\pm$0.12 & 98.44$\pm$0.03 &      & 94.75$\pm$0.10 & 95.98$\pm$0.10 & ***  \\
UMambaBot   & 98.54$\pm$0.07 & 98.55$\pm$0.06 &     & 98.52$\pm$0.05 & 98.84$\pm$0.02 &      & 95.69$\pm$0.04 & 96.48$\pm$0.09 & ***  \\
UMambaEnc   & 98.54$\pm$0.09 & 98.57$\pm$0.13 &     & 98.66$\pm$0.08 & 98.86$\pm$0.03 &      & 95.81$\pm$0.12 & 96.60$\pm$0.10 & ***  \\
SwinUMamba  & 98.80$\pm$0.03 & 98.86$\pm$0.02 & *   & 98.97$\pm$0.02 & 98.88$\pm$0.08 &      & 95.95$\pm$0.10 & 96.94$\pm$0.09 & ***  \\
\midrule
\multicolumn{10}{c}{\textbf{Foundation Models}}  \\
\midrule
SAM (box)     & 92.51$\pm$2.43 & 92.48$\pm$2.34 &     & 95.68$\pm$2.22 & 92.24$\pm$13.54 &     & 83.28$\pm$4.56 & 83.42$\pm$5.41 &     \\
SAM (pt)      & 92.25$\pm$5.13 & 91.95$\pm$7.04 &     & 97.55$\pm$2.37 & 96.18$\pm$6.02  &     & 83.16$\pm$5.49 & 80.58$\pm$15.15&     \\
MedSAM (box)  & 90.32$\pm$3.48 & 90.75$\pm$2.62 &     & 95.09$\pm$3.38 & 94.84$\pm$5.06  &     & 78.68$\pm$6.17 & 80.59$\pm$5.75 &     \\
\bottomrule
\end{tabular}}
\end{subtable}

\vspace{0.6cm}

\begin{subtable}{\textwidth}
\centering
\resizebox{\textwidth}{!}{
\begin{tabular}{l *{3}{ccc}}
\toprule
\multirow{2.5}{*}{\textbf{Model}} & 
\multicolumn{3}{c}{\textbf{Scaphoid}} & 
\multicolumn{3}{c}{\textbf{Trapezium}} & 
\multicolumn{3}{c}{\textbf{MC1}} \\
\cmidrule(lr){2-4}
\cmidrule(lr){5-7}
\cmidrule(lr){8-10}
 & \textbf{BE} & \textbf{nonBE} & \textbf{P} 
 & \textbf{BE} & \textbf{nonBE} & \textbf{P} 
 & \textbf{BE} & \textbf{nonBE} & \textbf{P} \\
\midrule
\multicolumn{10}{c}{\textbf{Supervised Models}}  \\
\midrule
UNet        & 96.86$\pm$0.15 & 96.77$\pm$0.12 &      & 95.30$\pm$0.20 & 95.84$\pm$0.21 & *    & 98.15$\pm$0.04 & 98.13$\pm$0.12 &      \\
DeepLabV3   & 96.56$\pm$0.10 & 96.73$\pm$0.06 &      & 94.88$\pm$0.13 & 95.78$\pm$0.08 & ***  & 97.91$\pm$0.03 & 98.01$\pm$0.02 & *    \\
FPN         & 96.22$\pm$0.14 & 96.54$\pm$0.12 & *    & 94.85$\pm$0.12 & 95.66$\pm$0.10 & ***  & 97.92$\pm$0.10 & 97.89$\pm$0.11 &      \\
PSPNet      & 94.35$\pm$0.44 & 94.82$\pm$0.25 & **   & 93.83$\pm$0.12 & 94.23$\pm$0.13 &      & 96.52$\pm$0.13 & 96.74$\pm$0.08 &      \\
DeepLabV3+  & 96.69$\pm$0.05 & 96.74$\pm$0.08 &      & 95.05$\pm$0.05 & 95.91$\pm$0.07 & ***  & 98.19$\pm$0.01 & 98.22$\pm$0.03 &      \\
SegResNet   & 96.54$\pm$0.31 & 96.62$\pm$0.21 &      & 94.91$\pm$0.32 & 95.54$\pm$0.30 & **   & 97.98$\pm$0.20 & 97.91$\pm$0.27 & *    \\
UNet++      & 97.37$\pm$0.06 & 97.24$\pm$0.06 & *    & 95.59$\pm$0.11 & 96.38$\pm$0.07 & ***  & 98.46$\pm$0.11 & 98.36$\pm$0.08 &      \\
SegFormer   & 96.66$\pm$0.15 & 96.91$\pm$0.06 &      & 95.19$\pm$0.07 & 95.96$\pm$0.07 & **   & 98.18$\pm$0.03 & 98.16$\pm$0.04 &      \\
TransUNet   & 97.73$\pm$0.03 & 97.73$\pm$0.03 &      & 96.03$\pm$0.08 & 96.74$\pm$0.05 & ***  & 98.68$\pm$0.03 & 98.62$\pm$0.07 & *    \\
UKAN        & 96.59$\pm$0.12 & 96.77$\pm$0.07 &      & 95.15$\pm$0.08 & 95.91$\pm$0.05 & **   & 98.23$\pm$0.05 & 98.13$\pm$0.09 & *    \\
UMambaBot   & 97.65$\pm$0.05 & 97.57$\pm$0.03 & *    & 95.81$\pm$0.07 & 96.65$\pm$0.09 & ***  & 98.63$\pm$0.03 & 98.54$\pm$0.05 & *    \\
UMambaEnc   & 97.62$\pm$0.13 & 97.59$\pm$0.11 &      & 95.87$\pm$0.12 & 96.75$\pm$0.06 & ***  & 98.68$\pm$0.03 & 98.57$\pm$0.11 & *    \\
SwinUMamba  & 97.92$\pm$0.04 & 97.82$\pm$0.03 &      & 96.17$\pm$0.08 & 96.87$\pm$0.03 & ***  & 98.76$\pm$0.04 & 98.74$\pm$0.05 &      \\
\midrule
\multicolumn{10}{c}{\textbf{Foundation Models}}  \\
\midrule
SAM (box)     & 87.82$\pm$5.20 & 87.58$\pm$5.07 &     & 82.26$\pm$8.43 & 84.52$\pm$6.37 &     & 97.17$\pm$1.08 & 97.12$\pm$1.06 &     \\
SAM (pt)      & 76.62$\pm$22.58 & 80.05$\pm$17.13 &     & 65.13$\pm$22.88 & 66.40$\pm$22.71 &     & 96.56$\pm$2.10 & 97.03$\pm$1.58 &     \\
MedSAM (box)  & 81.55$\pm$6.35 & 82.25$\pm$6.55 &     & 81.69$\pm$6.59 & 82.71$\pm$5.01 &     & 93.37$\pm$5.55 & 92.52$\pm$7.24 &     \\
\bottomrule
\end{tabular}}
\begin{tablenotes}
\item Foundation models: one inference (mean $\pm$ std across cases).\newline
Supervised models: five runs (mean $\pm$ std across runs).
\end{tablenotes}
\end{subtable}

\end{table*}

\begin{table*}[!t]
\centering
\caption{NSD performance on representative wrist bones. (Mann-Whitney U test between BE $\&$ nonBE, *: P $<$ 0.05; **: P $<$ 0.01; ***: P $<$ 0.001).}
\label{tab:joint_be_nonbe_nsd}

\begin{subtable}{\textwidth}
\centering
\resizebox{\textwidth}{!}{
\begin{tabular}{l *{3}{ccc}}
\toprule
\multirow{2.5}{*}{\textbf{Model}} & 
\multicolumn{3}{c}{\textbf{Radius}} & 
\multicolumn{3}{c}{\textbf{Ulna}} & 
\multicolumn{3}{c}{\textbf{Lunate}} \\
\cmidrule(lr){2-4}
\cmidrule(lr){5-7}
\cmidrule(lr){8-10}
 & \textbf{BE} & \textbf{nonBE} & \textbf{P} 
 & \textbf{BE} & \textbf{nonBE} & \textbf{P} 
 & \textbf{BE} & \textbf{nonBE} & \textbf{P} \\
\midrule
\multicolumn{10}{c}{\textbf{Supervised Models}}  \\
\midrule
Unet        & 80.57$\pm$0.62 & 82.78$\pm$0.51 & *   & 94.12$\pm$0.21 & 94.02$\pm$0.56 &     & 72.24$\pm$0.69 & 76.00$\pm$0.86 & *** \\
DeepLabV3   & 79.68$\pm$1.15 & 83.15$\pm$0.70 & *** & 93.01$\pm$0.43 & 93.40$\pm$0.21 &     & 71.71$\pm$1.13 & 75.45$\pm$0.57 & *** \\
FPN         & 78.61$\pm$0.91 & 82.12$\pm$0.43 & *** & 93.73$\pm$0.24 & 94.02$\pm$0.47 &     & 72.70$\pm$1.30 & 73.68$\pm$1.12 &     \\
PSPNet      & 69.94$\pm$1.87 & 70.40$\pm$1.80 &     & 83.78$\pm$2.46 & 82.90$\pm$1.97 &     & 61.34$\pm$1.35 & 61.63$\pm$1.10 &     \\
DeepLabV3+  & 81.31$\pm$0.45 & 84.27$\pm$0.37 & *** & 95.96$\pm$0.33 & 96.25$\pm$0.27 &     & 72.56$\pm$0.67 & 74.63$\pm$0.59 & *   \\
SegResNet   & 78.70$\pm$1.60 & 80.71$\pm$1.67 & *   & 94.07$\pm$1.06 & 93.86$\pm$0.20 &     & 71.80$\pm$1.58 & 74.15$\pm$1.80 & *   \\
Unet++      & 84.84$\pm$1.10 & 85.36$\pm$1.34 &     & 96.49$\pm$0.37 & 95.88$\pm$0.47 &     & 74.94$\pm$0.86 & 79.57$\pm$0.94 & *** \\
SegFormer   & 81.01$\pm$0.73 & 83.76$\pm$0.71 & *** & 95.72$\pm$0.35 & 95.78$\pm$0.29 &     & 74.84$\pm$1.33 & 77.14$\pm$0.64 & *   \\
TransUNet   & 87.96$\pm$0.72 & 89.63$\pm$0.72 & **  & 97.79$\pm$0.11 & 98.31$\pm$0.17 &     & 79.20$\pm$0.97 & 83.39$\pm$0.73 & *** \\
UKAN        & 79.61$\pm$2.39 & 81.90$\pm$2.60 & **  & 94.99$\pm$0.55 & 95.35$\pm$0.35 &     & 73.07$\pm$1.11 & 76.98$\pm$0.82 & *** \\
UMambaBot   & 86.83$\pm$1.42 & 87.33$\pm$0.88 &     & 96.66$\pm$0.29 & 97.95$\pm$0.12 & **  & 78.85$\pm$0.23 & 81.84$\pm$0.64 & *** \\
UMambaEnc   & 87.19$\pm$1.27 & 87.79$\pm$1.58 &     & 97.16$\pm$0.20 & 98.05$\pm$0.24 & *   & 79.42$\pm$0.86 & 82.34$\pm$0.72 & **  \\
SwinUMamba  & 90.63$\pm$0.50 & 90.95$\pm$0.35 &     & 98.45$\pm$0.21 & 98.15$\pm$0.36 &     & 81.05$\pm$0.58 & 84.81$\pm$0.60 & *** \\
\midrule
\multicolumn{10}{c}{\textbf{Foundation Models}}  \\
\midrule
SAM (box)     & 63.34$\pm$7.41  & 60.80$\pm$8.14  &     & 85.47$\pm$7.28  & 80.78$\pm$16.78 &     & 55.49$\pm$10.55 & 56.26$\pm$11.84 &     \\
SAM (pt)      & 63.67$\pm$11.79 & 62.22$\pm$13.06 &     & 93.45$\pm$8.22  & 91.00$\pm$13.38 &     & 54.76$\pm$13.71 & 54.01$\pm$17.61 &     \\
MedSAM (box)  & 44.22$\pm$16.33 & 45.73$\pm$11.62 &     & 74.88$\pm$18.11 & 75.06$\pm$19.63 &     & 32.76$\pm$15.99 & 40.32$\pm$16.99 & *   \\
\bottomrule
\end{tabular}}
\end{subtable}

\vspace{0.6cm}

\begin{subtable}{\textwidth}
\centering
\resizebox{\textwidth}{!}{
\begin{tabular}{l *{3}{ccc}}
\toprule
\multirow{2.5}{*}{\textbf{Model}} & 
\multicolumn{3}{c}{\textbf{Scaphoid}} & 
\multicolumn{3}{c}{\textbf{Trapezium}} & 
\multicolumn{3}{c}{\textbf{MC1}} \\
\cmidrule(lr){2-4}
\cmidrule(lr){5-7}
\cmidrule(lr){8-10}
 & \textbf{BE} & \textbf{nonBE} & \textbf{P} 
 & \textbf{BE} & \textbf{nonBE} & \textbf{P} 
 & \textbf{BE} & \textbf{nonBE} & \textbf{P} \\
\midrule
\multicolumn{10}{c}{\textbf{Supervised Models}}  \\
\midrule
Unet        & 84.69$\pm$1.41 & 82.52$\pm$1.14 & **  & 73.35$\pm$1.63 & 74.90$\pm$1.58 &     & 93.61$\pm$0.34 & 92.64$\pm$0.71 & **  \\
DeepLabV3   & 80.82$\pm$1.11 & 80.71$\pm$0.61 &     & 70.01$\pm$1.07 & 73.81$\pm$0.80 & **  & 92.29$\pm$0.33 & 92.24$\pm$0.16 &     \\
FPN         & 77.30$\pm$1.43 & 78.88$\pm$1.33 &     & 69.27$\pm$1.39 & 71.88$\pm$0.81 & *   & 91.02$\pm$0.70 & 90.10$\pm$0.81 & *   \\
PSPNet      & 61.72$\pm$2.67 & 63.81$\pm$1.73 &     & 60.33$\pm$1.06 & 59.37$\pm$1.37 &     & 81.94$\pm$1.14 & 81.31$\pm$0.44 &     \\
DeepLabV3+  & 81.97$\pm$0.51 & 81.71$\pm$1.06 &     & 71.63$\pm$0.46 & 74.73$\pm$0.75 & *   & 93.89$\pm$0.31 & 93.39$\pm$0.31 &     \\
SegResNet   & 80.83$\pm$2.88 & 81.20$\pm$1.61 &     & 70.73$\pm$2.73 & 72.48$\pm$2.10 &     & 92.39$\pm$0.91 & 91.14$\pm$1.64 & **  \\
Unet++      & 87.84$\pm$0.61 & 86.68$\pm$0.67 &     & 75.68$\pm$0.99 & 78.89$\pm$0.85 & *   & 95.06$\pm$0.77 & 94.19$\pm$0.72 &     \\
SegFormer   & 81.93$\pm$1.63 & 82.86$\pm$0.74 &     & 72.84$\pm$0.88 & 75.24$\pm$0.57 &     & 93.91$\pm$0.32 & 93.21$\pm$0.43 &     \\
TransUNet   & 90.86$\pm$0.27 & 91.01$\pm$0.45 &     & 79.43$\pm$0.43 & 82.38$\pm$0.37 & *   & 96.53$\pm$0.52 & 96.08$\pm$0.99 &     \\
UKAN        & 81.20$\pm$0.80 & 81.45$\pm$0.90 &     & 72.05$\pm$0.67 & 74.71$\pm$0.70 &     & 93.46$\pm$0.44 & 92.29$\pm$0.42 & *   \\
UMambaBot   & 90.60$\pm$0.38 & 89.75$\pm$0.09 &     & 78.82$\pm$0.33 & 82.13$\pm$0.90 & **  & 96.31$\pm$0.23 & 95.81$\pm$0.32 &     \\
UMambaEnc   & 90.56$\pm$0.97 & 89.86$\pm$0.70 &     & 79.46$\pm$0.70 & 83.09$\pm$0.55 & **  & 96.83$\pm$0.13 & 96.14$\pm$0.44 &     \\
SwinUMamba  & 92.47$\pm$0.36 & 91.80$\pm$0.13 &     & 81.13$\pm$0.53 & 84.02$\pm$0.17 & *   & 97.48$\pm$0.18 & 97.28$\pm$0.32 &     \\
\midrule
\multicolumn{10}{c}{\textbf{Foundation Models}}  \\
\midrule
SAM (box)     & 63.44$\pm$11.32 & 63.76$\pm$9.58  &     & 47.29$\pm$14.25 & 47.64$\pm$11.42 &     & 89.71$\pm$5.22  & 88.46$\pm$5.27  &     \\
SAM (pt)      & 51.73$\pm$25.27 & 54.21$\pm$21.42 &     & 33.50$\pm$21.40 & 34.66$\pm$19.81 &     & 88.06$\pm$7.41  & 88.70$\pm$6.37  &     \\
MedSAM (box)  & 26.40$\pm$14.98 & 26.92$\pm$11.81 &     & 35.81$\pm$10.66 & 31.37$\pm$9.96  & *   & 63.48$\pm$13.71 & 60.90$\pm$12.55 &     \\
\bottomrule
\end{tabular}}
\begin{tablenotes}
\item Foundation models: one inference (mean $\pm$ std across cases).\newline
Supervised models: five runs (mean $\pm$ std across runs).
\end{tablenotes}
\end{subtable}

\end{table*}

Table~\ref{tab:joint_be_nonbe_dsc} and Table~\ref{tab:joint_be_nonbe_nsd} report segmentation outcomes for representative wrist bones with and without BE. Both DSC and NSD indicate that joints affected by BE are harder to segment. The gap is generally small for DSC, typically about 1\%, but it is more evident for NSD, commonly 2\% to 6\%. Across both metrics, supervised methods outperform foundation models. Since BE alters bone morphology through erosion and deformation, analyzing BE versus nonBE groups allows us to assess whether such pathological changes affect segmentation accuracy.

For DSC (Table~\ref{tab:joint_be_nonbe_dsc}), supervised approaches achieve very high accuracy above 95\% in both the BE and nonBE groups. SwinUMamba attains the best scores on nearly all bones, remaining between 98\% and 99\% in nonBE cases and only slightly lower in BE cases, for example 98.0\% on the Distal Radius and 97.9\% on the Ulna. UMambaEnc and UMambaBot also maintain stable performance above 97\%. By contrast, CNN baselines such as UNet and DeepLabV3 yield slightly lower values, typically 96\% to 97\%, and foundation models perform worse overall with DSC around 90\% to 95\%, regardless of BE status.

For NSD (Table~\ref{tab:joint_be_nonbe_nsd}), the difference between BE and nonBE cases is more pronounced, especially in the Lunate and Trapezium. SwinUMamba again delivers the highest accuracy, reaching about 90\% to 97\% in nonBE cases and only a few percentage points lower in BE cases. UMambaEnc and UMambaBot remain competitive, whereas CNN baselines generally fall to the low 80s in BE joints. Foundation models show the weakest boundary accuracy, often dropping below 70\% in BE cases. For example, MedSAM records only 32.8\% on the Lunate with BE. These findings indicate that BE degrades boundary precision more than volumetric overlap, and that supervised Mamba family models, particularly SwinUMamba, are the most resilient to these challenges. This suggests that accurate segmentation in BE-affected regions remains a critical challenge for clinical applicability, as these areas are most relevant for disease monitoring and treatment decisions.

\clearpage

\subsubsection{Segmentation of Overlapping Regions}

\begin{table}[!t]
\centering
\caption{Instance segmentation results on overlapping regions. 
The best results in each column are highlighted in \textbf{bold}, 
and the second-best values are \underline{underlined}.}
\label{tab:overlap_mean}
\resizebox{\textwidth}{!}{
\begin{tabular}{p{2.5cm}>{\centering\arraybackslash}p{2.5cm}>{\centering\arraybackslash}p{2.5cm}>{\centering\arraybackslash}p{2.5cm}>{\centering\arraybackslash}p{2.5cm}>{\centering\arraybackslash}p{2.9cm}}
\toprule
\textbf{Model} & \textbf{DSC $\uparrow$ (\%)} & \textbf{NSD $\uparrow$ (\%)} & \textbf{VOE $\downarrow$ (\%)} & \textbf{MSD $\downarrow$ (pix)} & \textbf{MSD Fail Rate (\%)}\\
\midrule
\multicolumn{6}{c}{\textbf{Supervised Models}} \\
\midrule
UNet        & 65.80$\pm$0.83 & 66.60$\pm$1.30 & 46.10$\pm$0.88 & 2.77$\pm$0.15 & 3.28$\pm$0.31\\
DeepLabV3   & 68.60$\pm$0.36 & 67.01$\pm$0.45 & 43.49$\pm$0.36 & 2.27$\pm$0.04 & 3.07$\pm$0.10\\
FPN         & 67.11$\pm$0.39 & 64.68$\pm$0.82 & 45.32$\pm$0.45 & 2.41$\pm$0.03 & 2.83$\pm$0.20\\
PSPNet      & 61.12$\pm$0.54 & 54.42$\pm$0.60 & 52.07$\pm$0.55 & 3.10$\pm$0.07 & 4.20$\pm$0.15\\
DeepLabV3+  & 68.02$\pm$0.29 & 66.39$\pm$0.38 & 44.21$\pm$0.35 & 2.31$\pm$0.02 & 2.69$\pm$0.05\\
SegResNet   & 57.30$\pm$5.65 & 58.63$\pm$7.12 & 53.38$\pm$4.36 & 3.22$\pm$0.74 & 8.51$\pm$10.72\\
UNet++      & 70.16$\pm$0.68 & 71.31$\pm$0.57 & 41.46$\pm$0.62 & 2.16$\pm$0.04 & 2.81$\pm$0.21\\
SegFormer   & 68.61$\pm$0.15 & 67.94$\pm$0.35 & 43.37$\pm$0.20 & 2.30$\pm$0.04 & 2.63$\pm$0.13\\
TransUNet   & \underline{73.27$\pm$1.01} & \underline{75.66$\pm$0.51} & \underline{37.70$\pm$0.90} & \underline{1.91$\pm$0.04} & 2.61$\pm$0.23\\
UKAN        & 62.68$\pm$1.86 & 63.85$\pm$1.85 & 48.73$\pm$1.72 & 2.67$\pm$0.44 & 3.72$\pm$0.51\\
UMambaBot   & 72.70$\pm$0.18 & 74.55$\pm$0.40 & 38.55$\pm$0.17 & 2.08$\pm$0.18 & 2.60$\pm$0.11\\
UMambaEnc   & 72.45$\pm$0.47 & 74.67$\pm$0.51 & 38.79$\pm$0.52 & 1.97$\pm$0.03 & 2.77$\pm$0.15\\
SwinUMamba  & \textbf{74.45$\pm$0.25} & \textbf{77.15$\pm$0.20} & \textbf{36.25$\pm$0.27} & \textbf{1.83$\pm$0.02} & 2.69$\pm$0.20\\
\midrule
\multicolumn{6}{c}{\textbf{Foundation Models}} \\
\midrule
SAM (box)    & 3.78$\pm$2.83  & 2.51$\pm$1.88 & 97.09$\pm$2.34 & 9.49$\pm$7.79 & 89.96\\
SAM (pt)     & 3.41$\pm$1.90  & 2.58$\pm$1.72 & 97.91$\pm$1.29 & 49.63$\pm$23.80 & 61.13\\
MedSAM (box) & 5.32$\pm$4.16  & 3.34$\pm$2.52 & 96.31$\pm$3.12 & 12.10$\pm$5.43 & 75.79\\
\bottomrule
\end{tabular}}
\begin{tablenotes}
\item Foundation models: one inference (mean $\pm$ std across cases, MSD Fail Rate excluded). \newline
Supervised models: five runs (mean $\pm$ std across runs).
\end{tablenotes}
\end{table}

\begin{table*}[!t]
\centering
\caption{Overlap DSC performance on overlapping regions. 
The best results in each column are highlighted in \textbf{bold}, 
and the second-best values are \underline{underlined}.}
\label{tab:overlap_dsc}

\begin{subtable}{\textwidth}
\centering
\resizebox{\textwidth}{!}{
\begin{tabular}{lcccccccc}
\toprule
\textbf{Model} & \textbf{Cap-Sca} & \textbf{Cap-Tz} & \textbf{Cap-MC3} & \textbf{Radius-Lu} & \textbf{Radius-Sca} & \textbf{Ham-MC4} & \textbf{Ham-MC5} \\
\midrule
\multicolumn{8}{c}{\textbf{Supervised Models}} \\
\midrule
Unet        & 84.47$\pm$0.70 & 49.31$\pm$1.78 & 39.78$\pm$7.58 & 81.30$\pm$0.63 & 76.06$\pm$0.94 & 43.49$\pm$7.90 & 85.74$\pm$0.28 \\
DeepLabV3   & 84.85$\pm$0.26 & 50.78$\pm$2.23 & 49.65$\pm$0.70 & 82.73$\pm$0.73 & 76.18$\pm$0.88 & 51.81$\pm$1.82 & 85.86$\pm$0.20 \\
FPN         & 83.57$\pm$0.40 & 48.86$\pm$1.14 & 45.76$\pm$2.51 & 81.15$\pm$0.55 & 75.50$\pm$0.67 & 50.23$\pm$1.67 & 85.03$\pm$0.43 \\
PSPNet      & 75.76$\pm$0.74 & 49.85$\pm$1.10 & 35.00$\pm$1.90 & 73.82$\pm$0.44 & 69.53$\pm$1.68 & 39.91$\pm$1.21 & 80.24$\pm$0.54 \\
DeepLabV3+  & 84.04$\pm$0.47 & 50.46$\pm$0.59 & 49.34$\pm$1.83 & 82.75$\pm$0.12 & 76.73$\pm$0.15 & 51.92$\pm$1.86 & 85.48$\pm$0.18 \\
SegResNet   & 83.38$\pm$1.53 & 31.14$\pm$14.43 & 19.11$\pm$12.34 & 78.97$\pm$1.51 & 74.26$\pm$1.49 & 22.44$\pm$14.80 & 85.21$\pm$1.15 \\
Unet++      & 86.99$\pm$0.27 & 55.20$\pm$1.52 & 50.89$\pm$1.80 & 84.07$\pm$0.49 & 78.59$\pm$0.45 & 52.37$\pm$2.20 & 87.51$\pm$0.41 \\
SegFormer   & 85.12$\pm$0.39 & 49.64$\pm$0.86 & 49.08$\pm$1.20 & 83.03$\pm$0.54 & 77.59$\pm$0.83 & 50.73$\pm$0.98 & 86.34$\pm$0.23 \\
TransUNet   & \underline{88.96$\pm$0.19} & \underline{59.28$\pm$1.15} & 51.39$\pm$10.98 & \underline{88.04$\pm$0.26} & \underline{83.01$\pm$0.44} & \underline{57.01$\pm$1.23} & \underline{88.92$\pm$0.19} \\
UKAN        & 84.74$\pm$0.54 & 38.25$\pm$13.95 & 28.58$\pm$10.77 & 66.13$\pm$36.37 & 74.94$\pm$2.83 & 42.81$\pm$10.79 & 86.04$\pm$0.35 \\
UMambaBot   & 88.58$\pm$0.14 & 59.13$\pm$0.97 & \underline{56.24$\pm$0.91} & 85.74$\pm$0.64 & 81.01$\pm$0.60 & 56.31$\pm$1.06 & 88.48$\pm$0.16 \\
UMambaEnc   & 88.71$\pm$0.54 & 58.56$\pm$0.82 & 54.49$\pm$2.11 & 86.25$\pm$0.34 & 80.90$\pm$0.90 & 53.31$\pm$3.81 & 88.26$\pm$0.19 \\
SwinUMamba  & \textbf{89.71$\pm$0.13} & \textbf{60.01$\pm$0.52} & \textbf{57.51$\pm$0.64} & \textbf{88.09$\pm$0.13} & \textbf{83.40$\pm$0.27} & \textbf{59.59$\pm$0.78} & \textbf{89.48$\pm$0.19} \\
\midrule
\multicolumn{8}{c}{\textbf{Foundation Models}} \\
\midrule
SAM (box)    & 1.00$\pm$5.86 & 0.03$\pm$0.34 & 0.00$\pm$0.00 & 0.54$\pm$5.94 & 0.40$\pm$4.38 & 0.00$\pm$0.00 & 0.14$\pm$0.65 \\
SAM (pt)     & 2.09$\pm$7.30 & 0.69$\pm$4.10 & 0.12$\pm$0.54 & 0.84$\pm$6.05 & 0.66$\pm$5.06 & 0.31$\pm$1.13 & 1.21$\pm$3.53 \\
MedSAM (box) & 8.88$\pm$21.35 & 0.43$\pm$4.64 & 0.45$\pm$3.44 & 2.14$\pm$10.70 & 4.94$\pm$14.02 & 0.11$\pm$0.83 & 5.48$\pm$13.45 \\
\bottomrule
\end{tabular}}
\end{subtable}

\vspace{0.6cm}

\begin{subtable}{\textwidth}
\centering
\resizebox{\textwidth}{!}{
\begin{tabular}{lcccccccc}
\toprule
\textbf{Model} & \textbf{Lu-Sca} & \textbf{Sca-Tr} & \textbf{Tr-Tz} & \textbf{Tz-MC1} & \textbf{Tr-MC2} & \textbf{Tz-MC2} & \textbf{MC2-MC3} \\
\midrule
\multicolumn{8}{c}{\textbf{Supervised Models}} \\
\midrule
Unet        & 74.14$\pm$0.39 & 66.03$\pm$0.96 & 88.07$\pm$0.31 & 68.08$\pm$0.88 & 73.79$\pm$10.99 & 26.74$\pm$6.07 & 64.24$\pm$1.09 \\
DeepLabV3   & 73.72$\pm$0.50 & 66.20$\pm$0.72 & 88.08$\pm$0.20 & 69.26$\pm$0.28 & 80.35$\pm$0.42 & 33.18$\pm$0.83 & 67.74$\pm$0.49 \\
FPN         & 72.27$\pm$0.65 & 66.26$\pm$0.96 & 87.49$\pm$0.21 & 65.91$\pm$1.20 & 78.46$\pm$0.65 & 33.33$\pm$1.96 & 65.71$\pm$0.65 \\
PSPNet      & 63.95$\pm$1.10 & 61.35$\pm$0.39 & 85.16$\pm$0.30 & 58.42$\pm$1.39 & 72.12$\pm$0.61 & 31.25$\pm$1.58 & 59.39$\pm$1.66 \\
DeepLabV3+  & 72.65$\pm$0.43 & 66.50$\pm$0.43 & 88.01$\pm$0.14 & 66.68$\pm$0.92 & 79.07$\pm$0.59 & 32.02$\pm$1.14 & 66.62$\pm$0.77 \\
SegResNet   & 72.70$\pm$1.67 & 65.25$\pm$1.25 & 88.00$\pm$0.22 & 53.52$\pm$18.05 & 58.48$\pm$33.39 & 12.36$\pm$8.09 & 57.38$\pm$4.50 \\
Unet++      & 75.60$\pm$0.72 & 69.77$\pm$1.13 & 88.99$\pm$0.09 & 70.08$\pm$1.25 & 81.46$\pm$0.70 & 32.46$\pm$2.67 & 68.26$\pm$1.16 \\
SegFormer   & 73.72$\pm$0.79 & 67.87$\pm$0.73 & 88.39$\pm$0.17 & 70.48$\pm$0.64 & 80.39$\pm$0.32 & 31.28$\pm$0.74 & 66.85$\pm$0.34 \\
TransUNet   & 77.80$\pm$0.26 & \underline{72.16$\pm$0.50} & \underline{89.95$\pm$0.20} & \underline{74.58$\pm$0.58} & \underline{83.90$\pm$0.16} & \textbf{39.27$\pm$3.11} & \underline{71.50$\pm$0.59} \\
UKAN        & 73.87$\pm$0.83 & 65.82$\pm$1.42 & 88.39$\pm$0.17 & 63.17$\pm$3.31 & 79.20$\pm$1.10 & 25.24$\pm$3.39 & 60.31$\pm$7.51 \\
UMambaBot   & 77.67$\pm$0.33 & 71.11$\pm$0.65 & 89.44$\pm$0.15 & 74.14$\pm$0.85 & 83.14$\pm$0.32 & 36.43$\pm$0.77 & 70.32$\pm$0.55 \\
UMambaEnc   & \underline{77.75$\pm$0.35} & \textbf{72.17$\pm$0.51} & 89.79$\pm$0.19 & 73.97$\pm$0.44 & 82.90$\pm$0.43 & 36.79$\pm$1.79 & 70.38$\pm$0.77 \\
SwinUMamba  & \textbf{79.70$\pm$0.60} & 71.94$\pm$0.41 & \textbf{90.21$\pm$0.13} & \textbf{77.24$\pm$0.95} & \textbf{84.53$\pm$0.06} & \underline{38.71$\pm$1.21} & \textbf{72.18$\pm$0.68} \\
\midrule
\multicolumn{8}{c}{\textbf{Foundation Models}} \\
\midrule
SAM (box)    & 0.52$\pm$4.92 & 0.00$\pm$0.00 & 46.72$\pm$32.90 & 0.00$\pm$0.00 & 1.02$\pm$6.01 & 0.29$\pm$2.25 & 0.00$\pm$0.00 \\
SAM (pt)     & 1.14$\pm$3.49 & 1.24$\pm$4.44 & 31.73$\pm$17.32 & 0.05$\pm$0.31 & 3.85$\pm$5.32 & 1.93$\pm$6.69 & 0.10$\pm$0.53 \\
MedSAM (box) & 1.77$\pm$6.18 & 0.33$\pm$1.76 & 38.47$\pm$28.17 & 2.50$\pm$8.19 & 3.97$\pm$11.70 & 0.80$\pm$3.33 & 0.42$\pm$2.58 \\
\bottomrule
\end{tabular}}
\end{subtable}
\begin{tablenotes}
\item Foundation models: one inference (mean $\pm$ std across cases).\newline
Supervised models: five runs (mean $\pm$ std across runs).
\end{tablenotes}
\end{table*}

\begin{table*}[!t]
\centering
\caption{Overlap NSD performance on overlapping regions. 
The best results in each column are highlighted in \textbf{bold}, 
and the second-best values are \underline{underlined}.}
\label{tab:overlap_nsd}

\begin{subtable}{\textwidth}
\centering
\resizebox{\textwidth}{!}{
\begin{tabular}{lcccccccc}
\toprule
\textbf{Model} & \textbf{Cap-Sca} & \textbf{Cap-Tz} & \textbf{Cap-MC3} & \textbf{Radius-Lu} & \textbf{Radius-Sca} & \textbf{Ham-MC4} & \textbf{Ham-MC5} \\
\midrule
\multicolumn{8}{c}{\textbf{Supervised Models}} \\
\midrule
UNet        & 74.70$\pm$1.62 & 68.99$\pm$1.00 & 64.07$\pm$4.10 & 72.51$\pm$0.89 & 73.19$\pm$0.84 & 53.46$\pm$4.02 & 78.07$\pm$1.33 \\
DeepLabV3   & 72.88$\pm$0.76 & 67.01$\pm$1.18 & 67.08$\pm$0.65 & 71.83$\pm$1.73 & 70.15$\pm$1.77 & 57.32$\pm$0.46 & 77.14$\pm$0.32 \\
FPN         & 68.41$\pm$1.18 & 67.91$\pm$0.74 & 65.33$\pm$1.09 & 69.49$\pm$1.27 & 68.44$\pm$1.54 & 56.62$\pm$1.90 & 74.32$\pm$1.35 \\
PSPNet      & 51.44$\pm$1.21 & 62.87$\pm$0.98 & 55.30$\pm$0.83 & 56.47$\pm$0.98 & 57.62$\pm$2.00 & 44.42$\pm$0.59 & 61.76$\pm$1.50 \\
DeepLabV3+  & 71.16$\pm$1.40 & 67.90$\pm$1.31 & 67.72$\pm$0.75 & 71.37$\pm$0.62 & 70.40$\pm$0.47 & 58.40$\pm$0.58 & 75.64$\pm$0.59 \\
SegResNet   & 71.48$\pm$3.92 & 58.18$\pm$12.06 & 41.02$\pm$24.53 & 69.84$\pm$1.65 & 70.77$\pm$1.59 & 36.11$\pm$21.44 & 77.50$\pm$1.42 \\
UNet++      & 80.33$\pm$0.76 & 71.79$\pm$0.98 & 69.30$\pm$0.51 & 77.23$\pm$1.18 & 75.37$\pm$0.84 & 60.32$\pm$1.06 & 82.17$\pm$0.89 \\
SegFormer   & 73.21$\pm$1.41 & 69.25$\pm$0.75 & 67.99$\pm$0.62 & 71.79$\pm$0.81 & 71.10$\pm$1.89 & 59.48$\pm$0.92 & 79.04$\pm$0.71 \\
TransUNet   & \underline{84.69$\pm$0.57} & \underline{74.89$\pm$1.31} & 69.91$\pm$4.83 & \underline{84.13$\pm$0.16} & \underline{82.37$\pm$0.43} & \underline{66.49$\pm$1.64} & \underline{85.67$\pm$0.60} \\
UKAN        & 73.75$\pm$0.77 & 60.99$\pm$13.29 & 56.68$\pm$7.33 & 59.72$\pm$26.29 & 69.73$\pm$2.49 & 52.76$\pm$4.84 & 78.49$\pm$1.11 \\
UMambaBot   & 83.93$\pm$0.41 & \underline{74.89$\pm$0.31} & \underline{72.12$\pm$0.42} & 80.31$\pm$1.66 & 79.30$\pm$1.69 & 64.90$\pm$1.73 & 84.62$\pm$0.42 \\
UMambaEnc   & 84.32$\pm$1.41 & 74.88$\pm$1.06 & 71.52$\pm$0.59 & 81.15$\pm$1.32 & 79.16$\pm$1.40 & 62.80$\pm$3.55 & 84.25$\pm$0.35 \\
SwinUMamba  & \textbf{86.58$\pm$0.46} & \textbf{75.22$\pm$0.37} & \textbf{73.42$\pm$0.43} & \textbf{85.37$\pm$0.39} & \textbf{83.53$\pm$0.52} & \textbf{67.44$\pm$0.51} & \textbf{87.13$\pm$0.49} \\
\midrule
\multicolumn{8}{c}{\textbf{Foundation Models}} \\
\midrule
SAM (box)    & 0.98$\pm$5.41 & 0.16$\pm$1.73 & 0.00$\pm$0.00 & 0.26$\pm$2.84 & 0.47$\pm$4.71 & 0.00$\pm$0.00 & 1.31$\pm$5.52 \\
SAM (pt)     & 1.32$\pm$4.61 & 1.34$\pm$4.08 & 0.50$\pm$1.97 & 0.80$\pm$4.61 & 1.22$\pm$5.00 & 0.76$\pm$2.40 & 2.97$\pm$5.18 \\
MedSAM (box) & 3.64$\pm$8.67 & 0.17$\pm$1.89 & 0.73$\pm$5.19 & 2.14$\pm$8.44 & 4.62$\pm$10.37 & 0.15$\pm$1.52 & 6.18$\pm$12.19 \\
\bottomrule
\end{tabular}}
\end{subtable}

\vspace{0.6cm}

\begin{subtable}{\textwidth}
\centering
\resizebox{\textwidth}{!}{
\begin{tabular}{lcccccccc}
\toprule
\textbf{Model} & \textbf{Lu-Sca} & \textbf{Sca-Tr} & \textbf{Tr-Tz} & \textbf{Tz-MC1} & \textbf{Tr-MC2} & \textbf{Tz-MC2} & \textbf{MC2-MC3} \\
\midrule
\multicolumn{8}{c}{\textbf{Supervised Models}} \\
\midrule
UNet        & 68.15$\pm$0.60 & 68.29$\pm$0.93 & 63.58$\pm$1.52 & 72.96$\pm$1.32 & 57.88$\pm$10.60 & 50.17$\pm$5.65 & 66.45$\pm$1.39 \\
DeepLabV3   & 66.52$\pm$0.91 & 67.51$\pm$1.73 & 62.41$\pm$1.17 & 72.29$\pm$0.84 & 63.09$\pm$0.89 & 52.59$\pm$1.35 & 67.01$\pm$0.45 \\
FPN         & 62.54$\pm$1.43 & 67.31$\pm$0.62 & 59.06$\pm$1.30 & 67.25$\pm$1.30 & 58.59$\pm$1.86 & 52.31$\pm$1.63 & 64.68$\pm$0.82 \\
PSPNet      & 48.54$\pm$1.87 & 57.64$\pm$1.25 & 50.13$\pm$1.02 & 58.50$\pm$1.24 & 48.71$\pm$0.77 & 48.97$\pm$0.85 & 54.42$\pm$0.60 \\
DeepLabV3+  & 64.20$\pm$0.79 & 68.70$\pm$1.26 & 62.06$\pm$0.44 & 69.61$\pm$1.23 & 60.82$\pm$1.13 & 52.35$\pm$0.62 & 66.39$\pm$0.38 \\
SegResNet   & 65.16$\pm$2.72 & 67.23$\pm$1.39 & 63.21$\pm$1.64 & 59.98$\pm$14.16 & 45.19$\pm$26.18 & 31.04$\pm$18.28 & 58.63$\pm$7.12 \\
UNet++      & 71.29$\pm$1.61 & 72.70$\pm$1.66 & 67.06$\pm$0.48 & 75.67$\pm$1.43 & 67.84$\pm$1.59 & 55.13$\pm$1.63 & 71.30$\pm$0.57 \\
SegFormer   & 66.06$\pm$1.79 & 70.09$\pm$1.39 & 64.00$\pm$0.80 & 74.29$\pm$0.97 & 63.59$\pm$1.00 & 52.11$\pm$0.89 & 67.94$\pm$0.34 \\
TransUNet   & \underline{76.90$\pm$0.89} & 75.67$\pm$0.68 & \underline{71.88$\pm$0.80} & \underline{81.25$\pm$0.57} & \underline{73.38$\pm$0.35} & \textbf{58.16$\pm$2.20} & \underline{73.89$\pm$0.73} \\
UKAN        & 68.05$\pm$1.24 & 68.51$\pm$0.89 & 63.42$\pm$0.81 & 66.16$\pm$3.47 & 62.21$\pm$2.08 & 48.77$\pm$3.72 & 63.86$\pm$1.86 \\
UMambaBot   & 74.86$\pm$0.71 & 75.95$\pm$0.72 & 70.21$\pm$0.56 & 80.82$\pm$1.34 & 72.76$\pm$0.73 & 55.81$\pm$0.85 & 74.55$\pm$0.39 \\
UMambaEnc   & 75.08$\pm$1.18 & \underline{77.10$\pm$0.39} & 71.72$\pm$0.86 & 81.01$\pm$0.68 & 72.31$\pm$1.40 & 56.19$\pm$0.98 & \underline{74.67$\pm$0.51} \\
SwinUMamba  & \textbf{78.88$\pm$0.86} & \textbf{77.27$\pm$0.61} & \textbf{73.40$\pm$0.64} & \textbf{84.71$\pm$0.98} & \textbf{75.61$\pm$0.41} & \underline{56.37$\pm$0.83} & \textbf{75.24$\pm$0.64} \\
\midrule
\multicolumn{8}{c}{\textbf{Foundation Models}} \\
\midrule
SAM (box)    & 0.33$\pm$2.64 & 0.00$\pm$0.00 & 27.66$\pm$19.73 & 0.00$\pm$0.00 & 1.79$\pm$7.52 & 0.86$\pm$4.99 & 0.00$\pm$0.00 \\
SAM (pt)     & 1.67$\pm$4.55 & 1.85$\pm$5.79 & 12.41$\pm$12.48 & 0.45$\pm$1.77 & 5.24$\pm$8.63 & 4.72$\pm$9.65 & 0.36$\pm$1.65 \\
MedSAM (box) & 0.73$\pm$3.84 & 0.60$\pm$2.87 & 15.03$\pm$14.29 & 5.15$\pm$10.61 & 3.97$\pm$11.70 & 0.80$\pm$3.33 & 0.70$\pm$3.29 \\
\bottomrule
\end{tabular}}
\begin{tablenotes}
\item Foundation models: one inference (mean $\pm$ std across cases).\newline
Supervised models: five runs (mean $\pm$ std across runs).
\end{tablenotes}
\end{subtable}

\end{table*}

Although overall DSC and NSD values are high, visual inspection reveals that overlapping bones remain problematic, motivating a focused evaluation on these regions. Table~\ref{tab:overlap_mean}, Table~\ref{tab:overlap_dsc}, and Table~\ref{tab:overlap_nsd} demonstrate that overlapping wrist bones are particularly difficult to segment. SwinUMamba achieves the best performance across all metrics. In Table~\ref{tab:overlap_mean}, it reaches 74.5\% DSC and 77.2\% NSD, while also obtaining the lowest VOE (36.3\%) and the lowest MSD (1.83 pixels). The failure rate remains below 3\%, indicating strong robustness. In contrast, foundation models almost completely fail in this scenario, with DSC values below 6\% and NSD values below 4\%. These results show that such models cannot effectively separate closely packed structures without task-specific training.

For DSC in Table~\ref{tab:overlap_dsc}, SwinUMamba ranks first in nearly all pairwise regions. For example, it achieves 88.1\% on the Radius–Lunate interface and 89.7\% on the Capitate–Scaphoid interface. These results are about 1\% to 3\% higher than those of UMambaEnc and UMambaBot. CNN-based models such as UNet and DeepLabV3 are usually more than 10\% lower. TransUNet is the only Transformer baseline that approaches the Mamba-based models, with 86.3\% on Distal Radius–Lunate and 85.0\% on Trapezium–Meracarpal 2nd, but its performance is less consistent. SegResNet produces the lowest results, especially on Hamate-related overlaps, where its accuracy falls below 50\%.

For NSD in Table~\ref{tab:overlap_nsd}, a similar trend is observed, with boundary effects more evident. SwinUMamba again achieves the best results, reaching 87.1\% on the Hamate–MC5 interface and 86.6\% on the Capitate–Scaphoid interface, while remaining above 79\% even on difficult regions such as Trapezium–Trapezoid. UMambaEnc and UMambaBot follow closely, usually within 2\% to 3\% of SwinUMamba. TransUNet shows mixed performance; for instance, it improves to 69.9\% on Capitate–Metacarpal 3rd but lags behind on other pairs due to underestimation of overlap. CNN baselines drop further, often to 60\% to 70\%, and foundation models show the weakest performance, with NSD consistently below 5\% on almost all overlapping regions.

These results suggest that overlapping regions remain the most challenging aspect of wrist bone segmentation. Future work should focus on developing specialized strategies to improve performance in these areas, such as overlap-aware loss functions, boundary refinement modules, or targeted data augmentation. Enhancing segmentation accuracy in overlapping regions will be critical for achieving reliable and clinically applicable models. Improving overlap segmentation is especially important for clinical reliability, since diagnostic assessment often depends on accurate separation of adjacent bones in crowded anatomical areas.

\clearpage

\begin{figure}[!t]
    \centering
    \includegraphics[width=\linewidth]{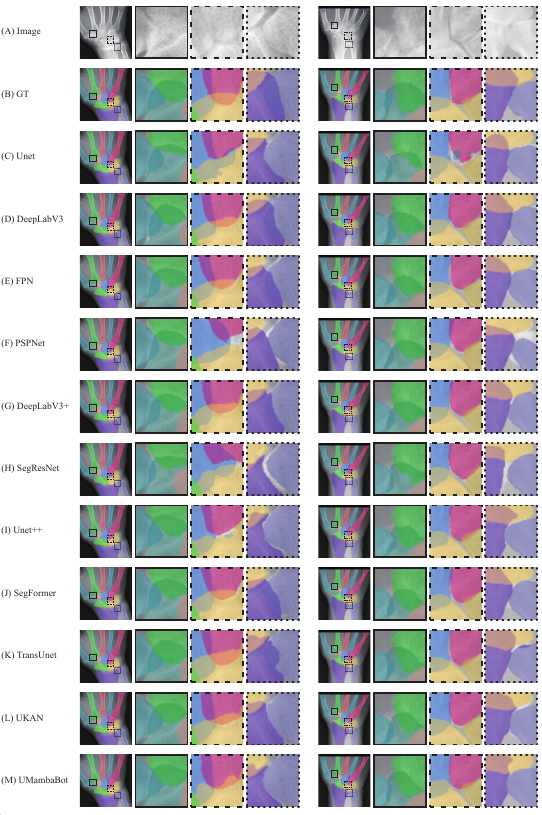}
    \caption{Additional visual results for wrist bone segmentation (Part A).}
    \label{fig:seg_results_AP_a}
\end{figure}
\begin{figure}[!t]
    \centering
    \includegraphics[width=\linewidth]{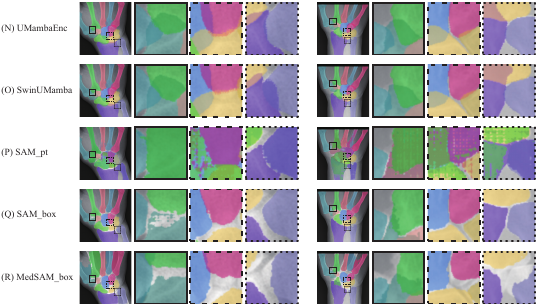}
    \caption{Additional visual results for wrist bone segmentation (Part B).}
    \label{fig:seg_results_AP_b}
\end{figure}

\subsubsection{Qualitative Visualization}
The visualization in Fig.~\ref{fig:seg_results_AP_a} and Fig.~\ref{fig:seg_results_AP_b} provides a detailed comparison of wrist bone segmentation results across different models. Visual inspection is crucial because numerical metrics alone may overlook local errors that are clinically significant, particularly around overlaps and pathological regions. All methods show noticeable errors compared to the ground truth, especially around boundaries where bones overlap or the image contrast is low. These problems are most pronounced in the Trapezium, Trapezoid, and Scaphoid, where irregular bone shapes and partial occlusions cause broken or incomplete predictions. In these regions, CNN-based models such as UNet, DeepLabV3, and FPN often produce blurred edges and fail to separate adjacent bones. Transformer-based models like TransUNet and SegFormer generate smoother boundaries, but they still lose fine details in overlapping areas. Mamba-based models, including UMambaEnc, UMambaBot, and SwinUMamba, show more stable performance. Their predictions follow bone contours more closely and reduce over-segmentation in crowded regions. SwinUMamba in particular achieves consistent delineation across both central and peripheral bones, with fewer gaps along thin boundaries. Nevertheless, even these models struggle in cases with severe overlap, where errors such as bone merging or missing edges remain visible. Foundation models perform poorly in visual comparison. SAM and MedSAM often produce coarse or fragmented masks that do not align with bone structures, highlighting their limitations when applied directly without fine-tuning. These models frequently miss small bones or collapse large bones into a single region, showing that task-specific supervision is essential for accurate wrist bone segmentation. In addition, BE-affected regions reveal another challenge. Across all models, bone erosion leads to irregular segmentation, with inward collapse or shape distortion often under-segmented. While some models occasionally capture these abnormalities, no architecture provides consistent results in such cases. This emphasizes the need for future work on methods that can better handle overlapping boundaries, subtle bone structures, and pathological deformations in order to achieve robust clinical applicability. Such qualitative evaluation further highlights that achieving clinically trustworthy segmentation requires not only high numerical scores but also consistent performance on challenging anatomical and pathological cases.

\subsubsection{Summary and Discussion}
Our quantitative and qualitative analyses lead to the following view. Current foundation models are not yet able to capture fine anatomical boundaries or resolve closely apposed bones in wrist radiographs, therefore supervised baselines remain necessary as clinically meaningful references that exploit pixel-level annotations. Although overall DSC and NSD are high, these global metrics can conceal systematic errors that concentrate at bone interfaces and in low-contrast zones; the overlap-focused evaluation exposes these failure modes and aligns with visual inspection. Bone erosion alters local geometry and degrades boundary fidelity more than volumetric overlap, hence BE-stratified reporting is essential for clinical relevance. Among supervised methods, Mamba-based architectures strike a favorable balance between global context and local detail, whereas Transformers may sacrifice small-structure precision and CNNs struggle at complex interfaces; this mechanism-level difference explains the observed ranking across DSC and NSD. Moving toward clinical reliability will likely require a combination of architectural and procedural advances, including interface-aware or contour-consistency losses, boundary refinement and instance disambiguation modules, targeted augmentation that simulates occlusion and erosion, sampling curricula that oversample rare overlap patterns and BE cases, active learning to prioritize uncertain regions for annotation, and uncertainty estimation or test-time adaptation to mitigate distribution shift. Evaluation practice should likewise move beyond single numbers by reporting per-interface metrics, BE-stratified results, failure rates, and distance-based errors alongside DSC and NSD. Together, these directions align the benchmark with clinical priorities and outline a path toward robust and deployable wrist bone segmentation.

\begin{figure}[!t]
    \centering
    \includegraphics[width=\linewidth]{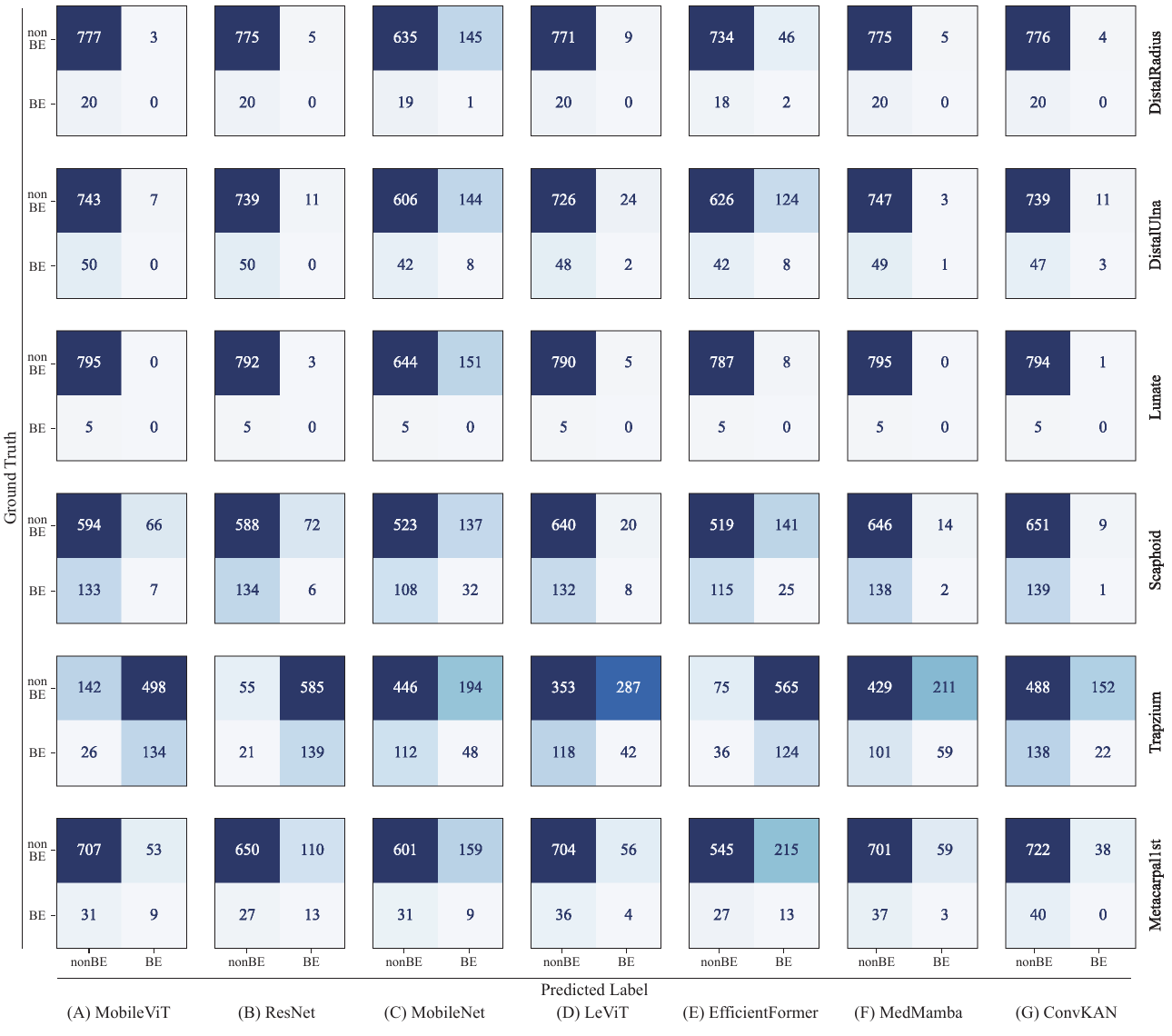}
    \caption{Confusion matrix results for classification of BE and nonBE.}
    \label{fig:be_results_AP}
\end{figure}

\subsection{Classification of BE}
Figure~\ref{fig:be_results_AP} shows confusion matrices for BE and nonBE classification across six representative carpal joints. Most models achieve high accuracy on nonBE cases, where predictions are strongly concentrated along the diagonal. However, substantial misclassification occurs in BE cases, reflecting the inherent difficulty of detecting pathological changes. This contrast shows that overall accuracy can be misleading, as it is dominated by the abundance of nonBE cases while failing to reflect systematic errors in BE detection. For example, MobileNet and ResNet frequently misclassify BE samples as nonBE in the Distal Ulna and Trapezium, indicating a bias toward conservative predictions. Similar trends are observed in the Scaphoid, where BE cases are often confused with nonBE due to irregular joint boundaries. This joint-specific variability indicates that anatomical complexity directly affects classification difficulty, and a single model may not perform equally well across all regions.

More advanced architectures show partial improvements. MedMamba and ConvKAN achieve more balanced predictions, particularly in the Lunate and Distal Radius, where BE cases are identified with higher sensitivity compared to earlier CNN-based models. Nevertheless, even these models still exhibit notable false negatives, especially in challenging regions such as the Scaphoid and Trapezium. This suggests that while recent methods better capture morphological changes, robust recognition of erosive patterns remains unresolved. Clinically, missing even a small number of BE cases may delay diagnosis or underestimate disease severity, underscoring the need for higher sensitivity in BE detection. These results highlight the importance of designing models capable of learning discriminative features that generalize well to pathological variations, especially in early-stage RA where accurate BE detection is clinically critical.

Future work should address the extreme class imbalance between BE and nonBE cases through techniques such as focal loss or targeted data augmentation, and explore generative approaches for synthesizing BE-like patterns, in order to enhance the ability of models to capture subtle pathological features.

\section{Broader Impact}
This work provides a publicly available and well-annotated multi-task wrist dataset and benchmark designed to advance research in RA diagnosis using conventional wrist radiographs. This resource enables researchers to build and evaluate advanced models for RA-related tasks with consistency and rigor. The authors do not anticipate any negative societal impacts stemming from this work. On the contrary, a positive impact may arise through the development of robust computer-aided diagnosis systems, which can facilitate early detection and monitoring of RA with reduced reliance on manual annotations. This has the potential to enhance clinical efficiency, reduce expert workload, and improve access to specialized care, particularly in under-resourced healthcare settings.

\bibliographystyle{plain}
\bibliography{refs}

\end{document}